\documentclass[rsi,amsmath,amssymb,floatfix,longbibliography,reprint,superscriptaddress]{revtex4-1}
\usepackage{times} 
\usepackage{graphicx}
\usepackage{etoolbox}
\usepackage{dcolumn}
\usepackage{xcolor}
\usepackage{underscore}
\usepackage{multirow}
\usepackage{float}
\usepackage{placeins}
\usepackage[T1]{fontenc}
\usepackage[utf8]{inputenc}

\usepackage{subcaption}

\graphicspath{{./figs/}}

\makeatletter
\def\@email#1#2{%
 \endgroup
 \patchcmd{\titleblock@produce}
  {\frontmatter@RRAPformat}
  {\frontmatter@RRAPformat{\produce@RRAP{*#1\href{mailto:#2}{#2}}}\frontmatter@RRAPformat}
  {}{}
}%
\makeatother

\begin{document}

\title{Automated tomographic assessment of structural defects of freeze-dried pharmaceuticals}

\author{Patric Müller}
\affiliation{Institut für Multiskalensimulation, Friedrich-Alexander-Universität Erlangen-Nürnberg, Germany}
\author{Achim Sack}
\affiliation{Institut für Multiskalensimulation, Friedrich-Alexander-Universität Erlangen-Nürnberg, Germany}
\author{Jens Dümler}
\affiliation{Institut für Multiskalensimulation, Friedrich-Alexander-Universität Erlangen-Nürnberg, Germany}
\author{Michael Heckel}
\affiliation{Institut für Multiskalensimulation, Friedrich-Alexander-Universität Erlangen-Nürnberg, Germany}
\affiliation{IT Unit, University of Technology Nuremberg, Germany}
\author{Tim Wenzel}
\affiliation{Division of Pharmaceutics, Freeze Drying Focus Group, Friedrich-Alexander-Universität Erlangen-Nürnberg, Germany}
\affiliation{GILYOS GmbH, Würzburg, Germany}
\author{Teresa Siegert}
\affiliation{Division of Pharmaceutics, Freeze Drying Focus Group, Friedrich-Alexander-Universität Erlangen-Nürnberg, Germany}
\author{Sonja Schuldt-Lieb}
\affiliation{medac GmbH, Wedel, Germany}
\author{Henning Gieseler}
\affiliation{GILYOS GmbH, Würzburg, Germany}
\author{Thorsten Pöschel}
\affiliation{Institut für Multiskalensimulation, Friedrich-Alexander-Universität Erlangen-Nürnberg, Germany}
\date{\today}

\begin{abstract}
The topology and surface characteristics of lyophilisates significantly impact the stability and reconstitutability of freeze-dried pharmaceuticals. Consequently, visual quality control of the product is imperative. However, this procedure is not only time-consuming and labor-intensive but also expensive and prone to errors. In this paper, we present an approach for fully automated, non-destructive inspection of freeze-dried pharmaceuticals, leveraging robotics, computed tomography, and machine learning.
\end{abstract}
\maketitle

\section{Introduction - structural defects in freeze-dried pharmaceuticals}
\label{sec:introduction}

Freeze-drying (lyophilization) enables the heat-free removal of water from products through sublimation. This process substantially improves the stability of various pharmaceutical products \cite{pikal:2010,wang:2000,Hardter2023CGPT,Pardeshi2023CGPT}. For many biopharmaceutical products, particularly those based on therapeutic proteins, lyophilization is essential for ensuring stability \cite{Manning:2010,Rebizzi2023CGPT,Carpenter2020LyophilizationOPCGPT,Remmele2012DevelopmentOSCGPT}. The resulting products, sealed in glass vials, can be conveniently handled and later reconstituted to their original form, for example, for injection through rehydration.

The pore morphology of freeze-dried products is directly influenced by the formulation components and process parameters, playing a crucial role in drying performance and product quality attributes \cite{pikal:2007}. In a typical freeze-drying cycle, the primary drying step is optimized to maintain a minimal safety margin below the critical temperature of the formulation. This optimization aims to achieve maximum drying efficiency while preserving the desired pore morphology \cite{tang:2004}. Product defects, such as the collapse of the drying matrix, may result from temperatures surpassing the critical formulation temperature \cite{Liu2005StudyOTCGPT,Butreddy2020LyophilizationOSCGPT}. This is of significance for product quality as it can lead to issues such as water entrapment, reduced specific surface areas, and higher residual moisture contents \cite{tang:2004,barresi:2009}.

\begin{figure*}[ht!]
    \begin{subfigure}{\textwidth}
        \includegraphics[width=0.1125\columnwidth]{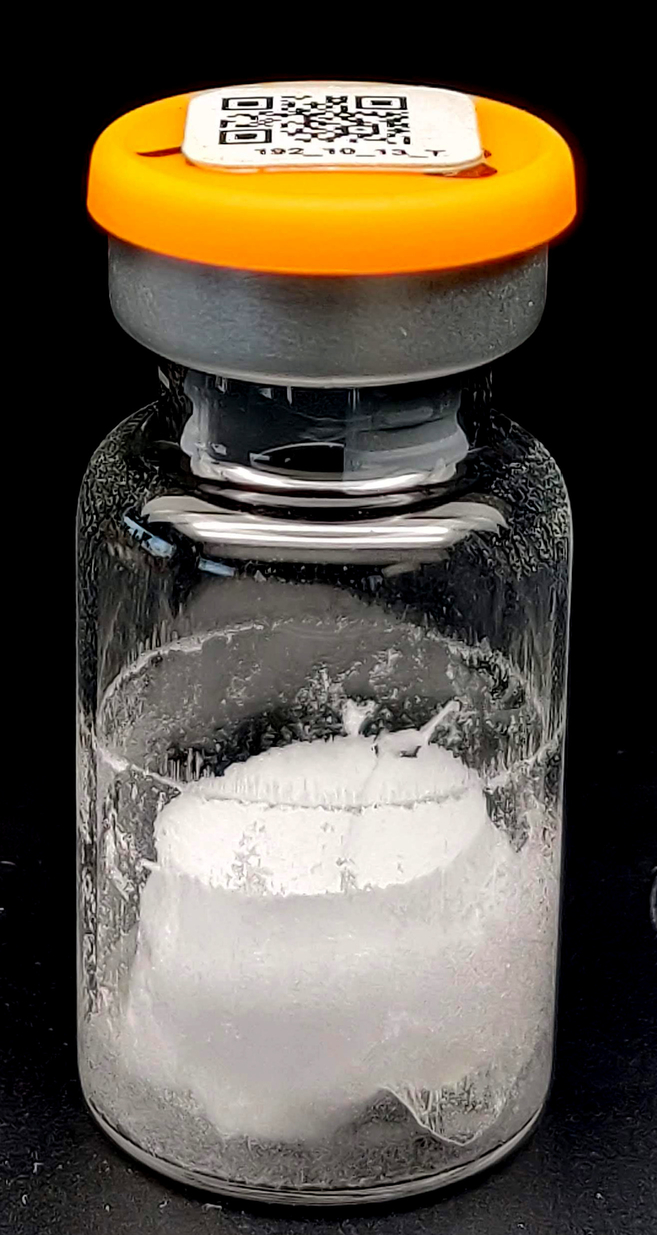} \hspace*{1cm} \includegraphics[width=0.2\columnwidth]{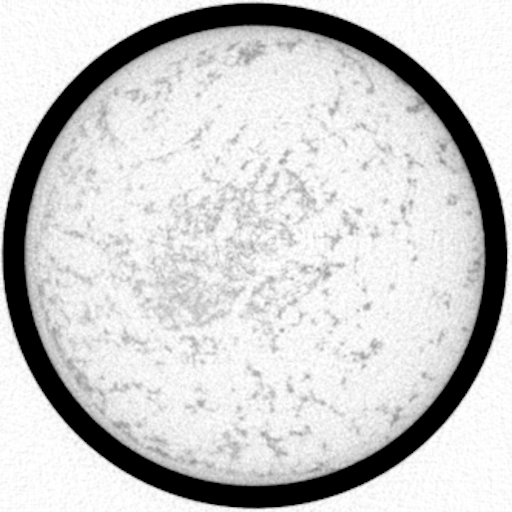}\hspace*{1cm}
        \centering\includegraphics[width=0.5\columnwidth]{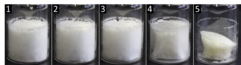}\hspace*{\fill}
        \caption{shrinkage / collapse}
        \label{fig:defects:shrink}
    \end{subfigure}
    \begin{subfigure}{\textwidth}
        \includegraphics[width=0.1125\columnwidth]{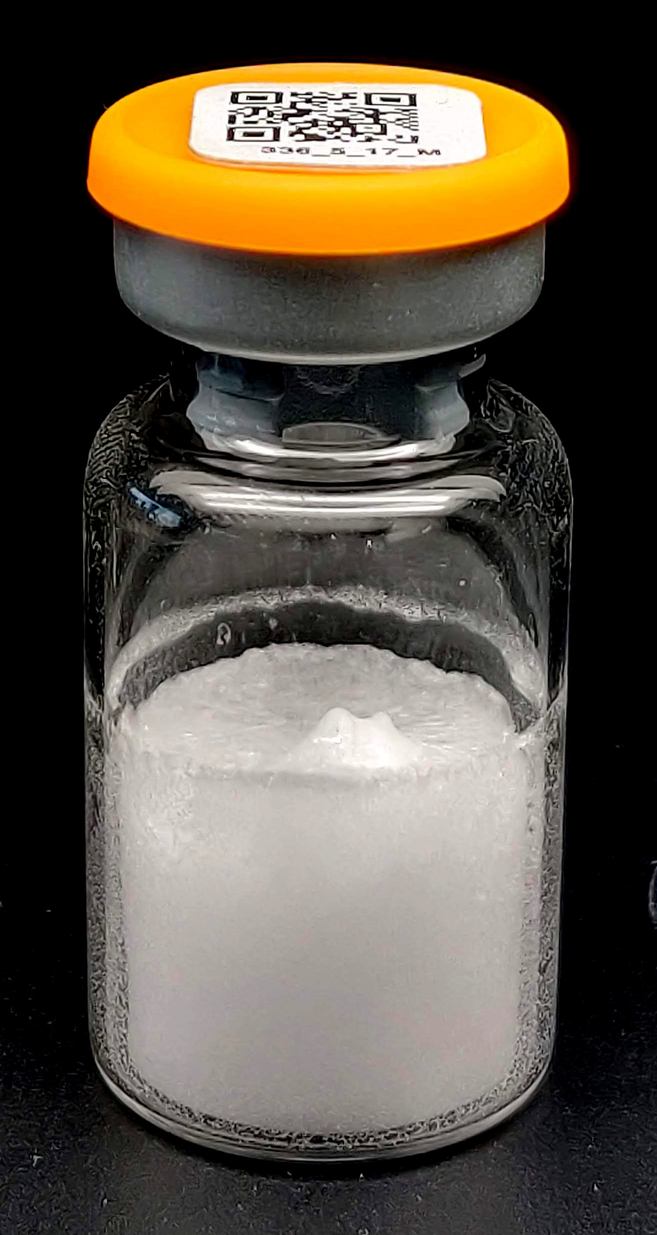} \hspace*{1cm} \includegraphics[width=0.2\columnwidth]{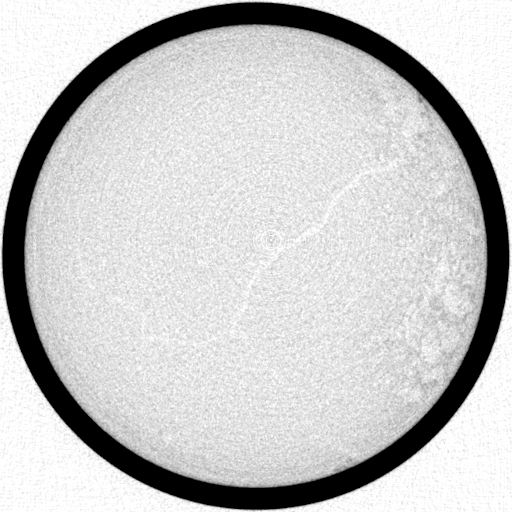}\hspace*{1cm}
        \centering\includegraphics[width=0.225\columnwidth]{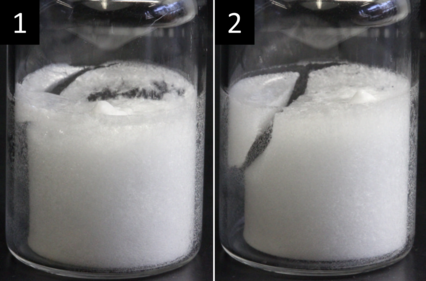}\hspace*{\fill}
        \caption{crack formation}
        \label{fig:defects:crack}
    \end{subfigure}
    \begin{subfigure}{\textwidth}
        \includegraphics[width=0.1125\columnwidth]{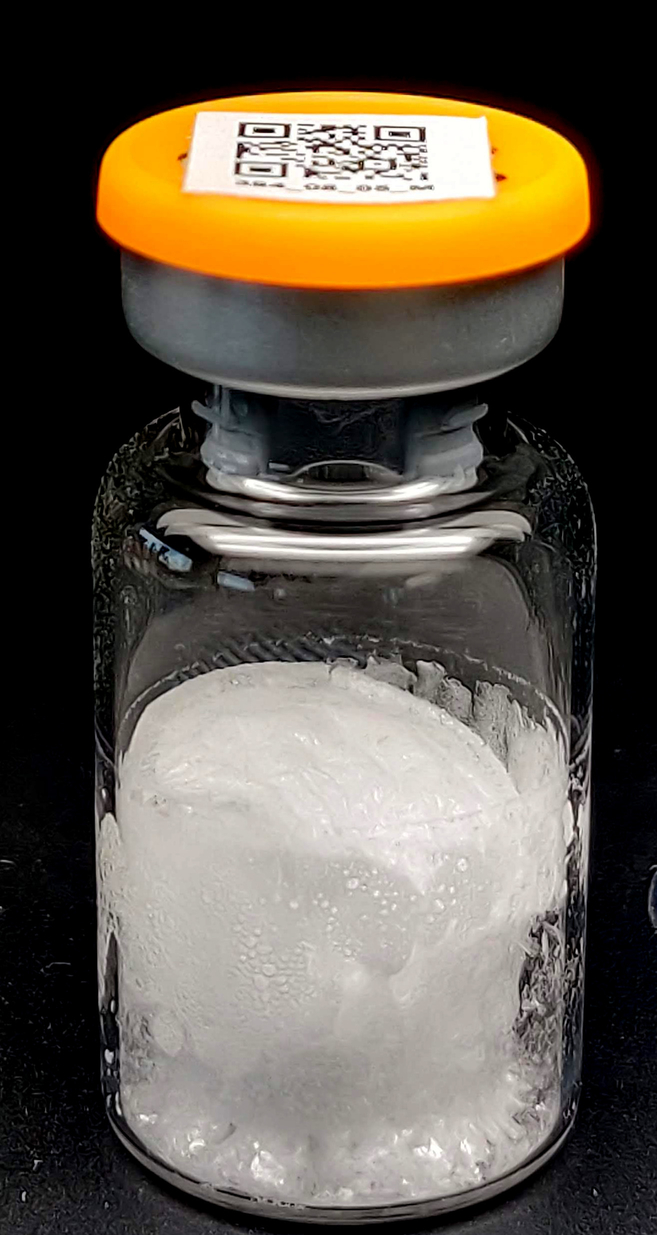} \hspace*{1cm} \includegraphics[width=0.2\columnwidth]{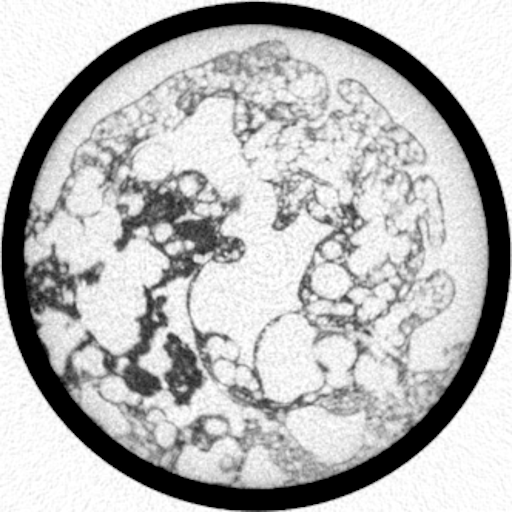}\hspace*{1cm}
        \centering\includegraphics[width=0.225\columnwidth]{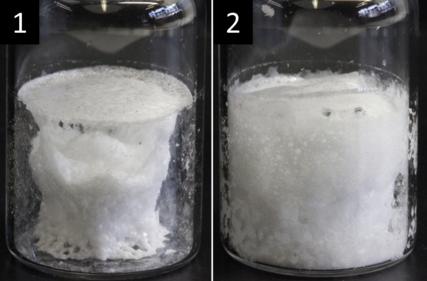}\hspace*{\fill}
        \caption{foaming/ bubble formation}
        \label{fig:defects:foam}
    \end{subfigure}
    \begin{subfigure}{\textwidth}
        \includegraphics[width=0.1125\columnwidth]{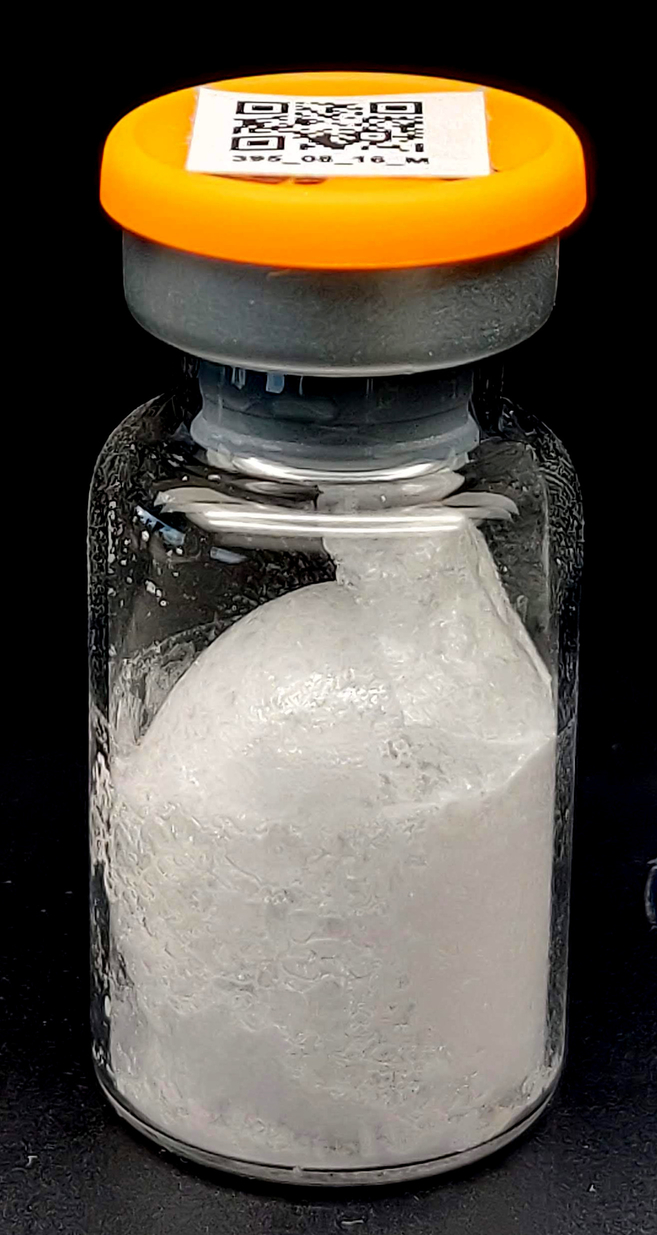} \hspace*{1cm} \includegraphics[width=0.2\columnwidth]{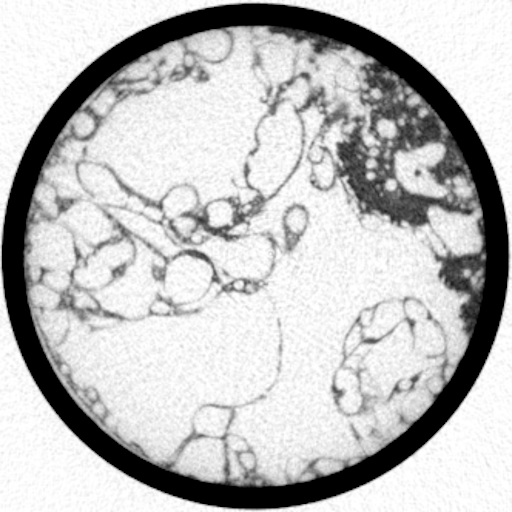}\hspace*{1cm}
        \centering\includegraphics[width=0.225\columnwidth]{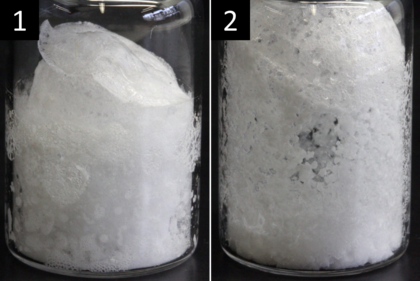}\hspace*{\fill}
        \caption{blow-out}
        \label{fig:defects:blowout}
    \end{subfigure}
    \begin{subfigure}{\textwidth}
        \includegraphics[width=0.1125\columnwidth]{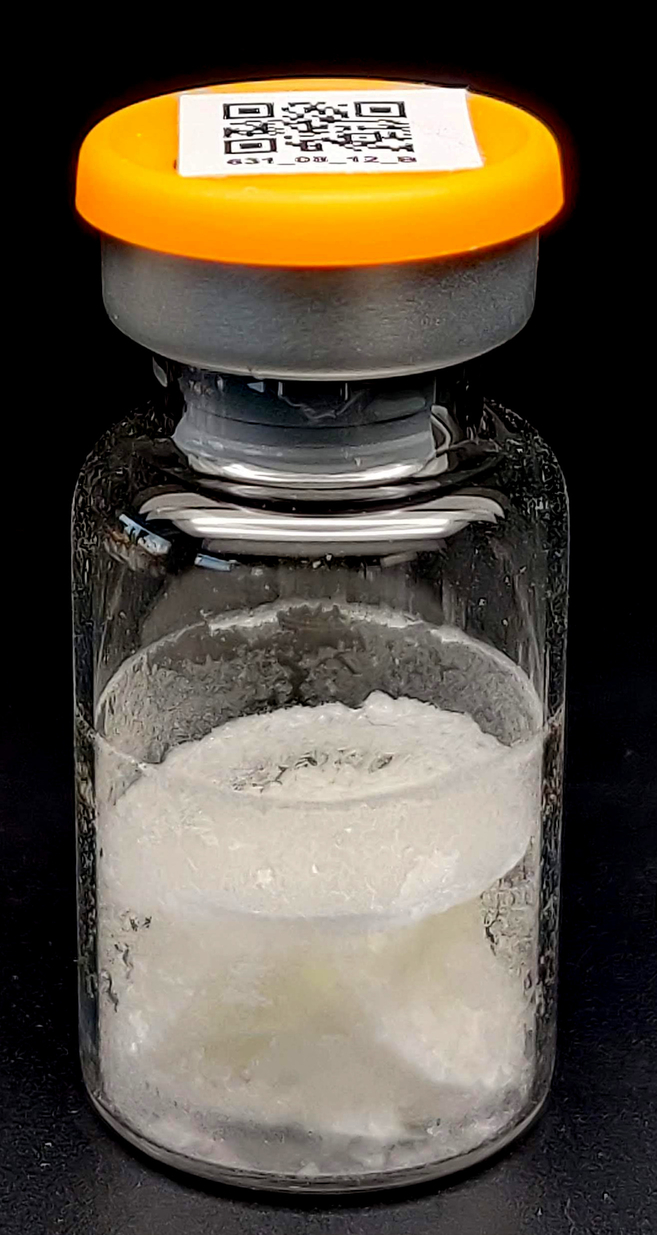} \hspace*{1cm} \includegraphics[width=0.2\columnwidth]{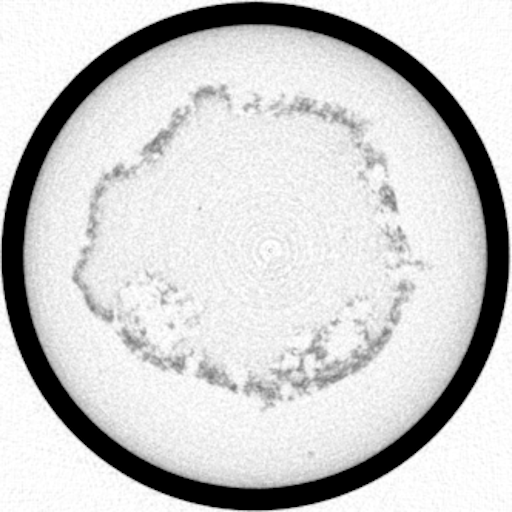}\hspace*{1cm}
        \centering\includegraphics[width=0.225\columnwidth]{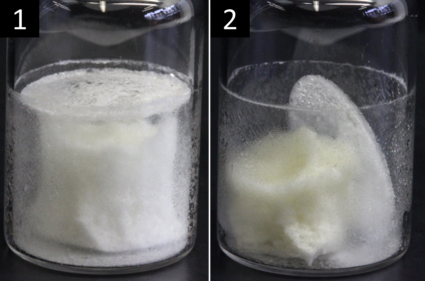}\hspace*{\fill}
        \caption{loose-skin}
        \label{fig:defects:skin}
    \end{subfigure}
    \caption{Each line (subfigure) depicts one of the typical morphological defects in freeze-dried pharmaceuticals as listed in Table \ref{tab:defect}. The first column of each line presents a characteristic image of the respective defect, the second column displays a tomographic vertical slice of the lyophilisate, and the third column illustrates the various degrees of severity associated with the corresponding defect. \label{fig:defects}}
\end{figure*}

In line with industry guidelines, correct volume and cake appearance are cited as critical attributes during the visual inspection of freeze-dried products \cite{inspection:2014}. A recent commentary provides an overview of macroscopical product appearances and their potential impact on product quality attributes for freeze-dried products \cite{patel:2017}. Various morphological defects, each with varying degrees of severity as defined in Table \ref{tab:defect}, are listed below. Example images illustrating each defect and its degrees of severity are presented in Figure \ref{fig:defects}.
\begin{table}[ht]
\caption{Criteria for characterization by optical inspection}
\label{tab:defect} 
\begin{tabular}{| p{0.3\columnwidth} |c| p{0.54\columnwidth} |}
\hline
\textbf{macroscopic trait} & \textbf{rating} & \textbf{criteria for rating}\\
\hline
\multirow{7}{*}{\parbox[t]{0.3\columnwidth}{shrinkage /\\ collapse}} & 1 & very little to no shrinkage\\ \cline{2-3}
 & 2 & minor local shrinkage\\ \cline{2-3}
 & 3 & overall shrinkage, macroscopic shape intact\\ \cline{2-3}
 & 4 & severe shrinkage, local macroscopic collapse\\ \cline{2-3}
 & 5 & severe shrinkage, overall collapse\\ \hline
\multirow{3}{*}{\parbox[t]{0.3\columnwidth}{Crack formation}} & 0 & none\\ \cline{2-3}
 & 1 & minor \\ \cline{2-3}
 & 2 & major\\ \hline
\multirow{3}{*}{\parbox[t]{0.3\columnwidth}{foaming /\\ bubble formation}} & 0 & none\\ \cline{2-3}
 & 1 & minor\\ \cline{2-3}
 & 2 & major\\ \hline
\multirow{5}{*}{\parbox[t]{0.3\columnwidth}{blow-out /\\ rising of cake structure}} & 0 & none\\ \cline{2-3}
 & 1 & minor, product not in contact with stopper\\ \cline{2-3}
 & 2 & major, product in contact with stopper\\ \hline
\multirow{3}{*}{\parbox[t]{0.3\columnwidth}{loose skin}} & 0 & none\\ \cline{2-3}
 & 1 & partially loose skin\\ \cline{2-3}
 & 2 & completely loose skin\\
\hline
\end{tabular}
\end{table}

In this study, we investigate several morphological defects observed during the freeze-drying process of pharmaceuticals:
\begin{itemize}
\item \textit{Shrinkage/Collapse:} The lyophilisate undergoes shrinkage, and in severe cases, it detaches from the vial wall. While some shrinkage is an anticipated outcome, more severe manifestations may indicate an inadequately controlled drying process and residual moisture in the sample.

\item \textit{Crack Formation:} Large stresses generated during the drying process cause the sample to break into multiple pieces.

\item \textit{Foaming/Bubble Formation:} Reduced pressure and applied heat may induce the formulation to boil, resulting in foaming during the process. Depending on the formulation's ability to stabilize the gas-liquid surface, this may lead to the presence of large gas bubbles in the lyophilisate, alongside dense filaments of dried product.

\item \textit{Blow Out/Rising of Cake Structure:} Connected to bubble formation, this effect focuses on the increase in the sample's volume. In severe cases, liquid sample is forced through the vial opening and lost.

\item \textit{Loose Skin:} Evaporation occurring at the fluid surface leads to the formation of a solid skin or crust. Depending on formulation and process parameters, the skin can either form a cap or subsequently detach from the dried product.
\end{itemize}

As structural defects significantly impact the lyophilisate quality, freeze-dried products undergo routine visual inspections \cite{fda:2014,rambhatla:2004,schersch:2010,liu:2006,kim:1998,wang:2000,pikal:2010}. Additional inspections involve sporadic checks of microstructure and pore structure using destructive analytical techniques such as SEM \cite{startzel:2015b} or SSA \cite{geidobler:2013}. However, these destructive techniques are impractical for high-throughput screening on an industrial scale. Human visual inspection is time-consuming, expensive, and tiring for workers, making it error-prone. Automated camera systems with downstream image analysis are already in use \cite{tsay:2019}. While these automated methods reduce the need for human interaction, they are limited to examining surface properties of the lyophilisate.

Quality-relevant defects, such as the formation of cracks, can also occur inside the product, hindering visual inspection. In this study, we propose quality control using X-ray computed tomography (CT) for a volumetric assessment of freeze-dried products. Although CT has been manually applied to investigate individual freeze-dried products \cite{haeuser:2018,pisano:2017,kunz:2019}, we introduce an apparatus that eliminates the need for human intervention through the use of robotics and artificial intelligence. While the system presented is a conceptual study not immediately suitable for industrial use, it serves as a starting point for potential future high-throughput processes.

CT was employed to analyze defects in tablets \cite{schomberg:2021,yost:2019,sondej:2015,zeitler:2009,hancock:2005} and powders used in dry powder inhalers \cite{gajjar:2020}. An essential application of freeze-drying is to safeguard protein-based pharmaceutical products against damage during transport and handling. The question of whether such substances withstand exposure to X-rays was addressed in \cite{wenzel:2021}, where freeze-dried biopharmaceuticals were irradiated with approximately 100 Gy. No adverse effects on the chemical and physical stability of three model formulations from different substance classes were observed. Consequently, CT can be considered a valuable tool for non-destructive quality control of freeze-dried pharmaceuticals. Exemplary tomograms of lyophilisates are presented in the second column of Fig. \ref{fig:defects}.

\section{Sketch of the automated X-ray setup}
\label{sec:hardware}

\begin{figure}[ht]
\centerline{\includegraphics[width=\columnwidth]{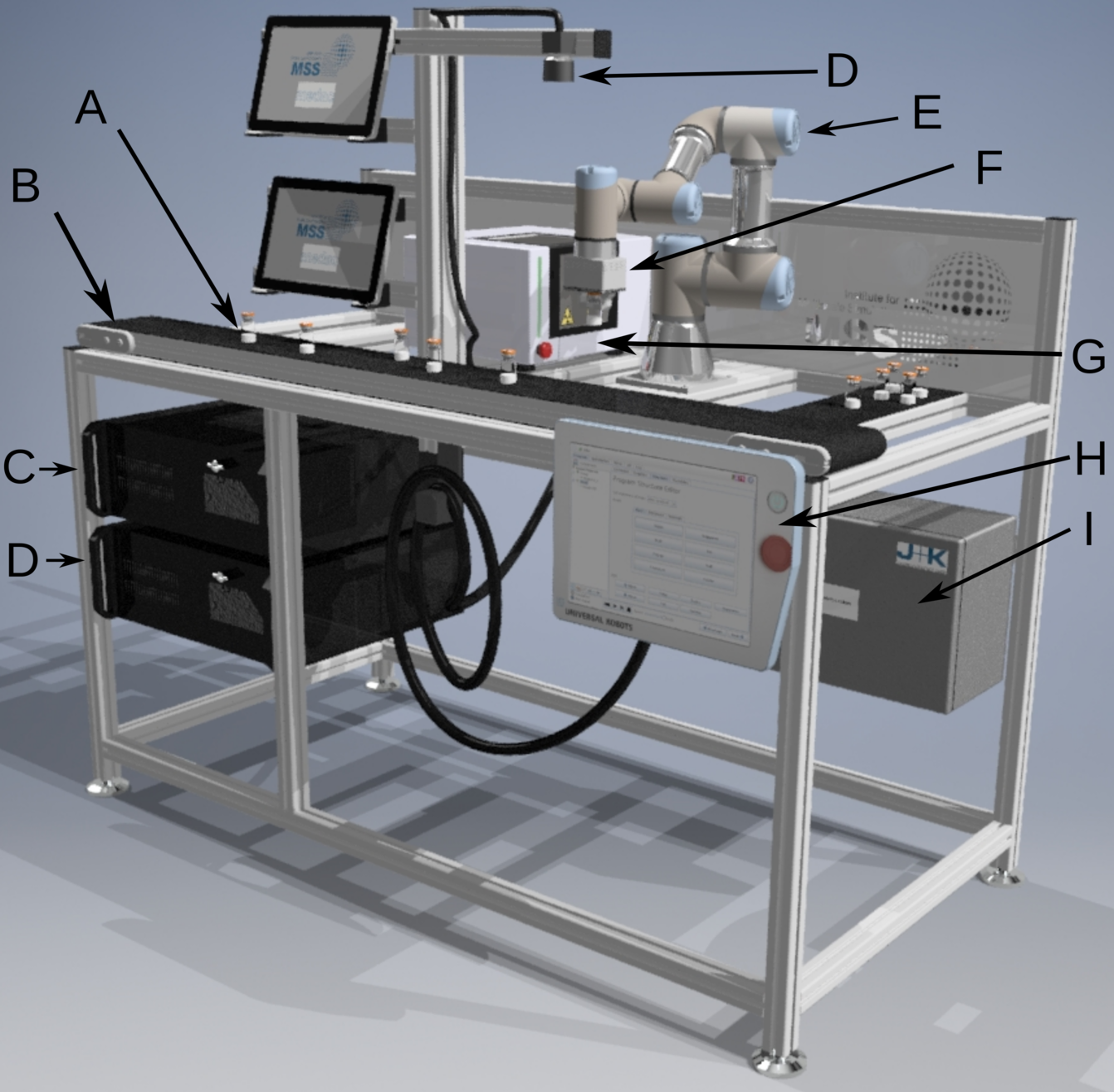}}
\caption{
Apparatus sketch depicting key components: (A) vials, (B) conveyor belt, (C) PC for CT, (D) PC for robot control and image analysis, (E) camera, (F) robotic arm, (G) X-ray tomograph, (H) teach pendant, (I) robot controller.}
\label{fig:cad}
\end{figure}

The fully automatic tomographic control of the structural quality of freeze-dried pharmaceuticals encompasses the following tasks:
\begin{itemize}
 \item\label{item:hardware:feed} automatic feeding of the samples contained in standard vials
 \item\label{item:hardware:id} detection and identification of the vials through individual QR codes
 \item\label{item:hardware:moveIn} placement of the samples into the tomograph
 \item\label{item:hardware:ct} recording of the tomogram
 \item\label{item:hardware:moveOut} removal of the sample from the tomograph
 \item reconstruction of the CT-data
 \item classification of the sample
 \item\label{item:hardware:sort} output positioning based on the classification
\end{itemize}

\begin{figure}[ht]
\centerline{\includegraphics[width=\columnwidth]{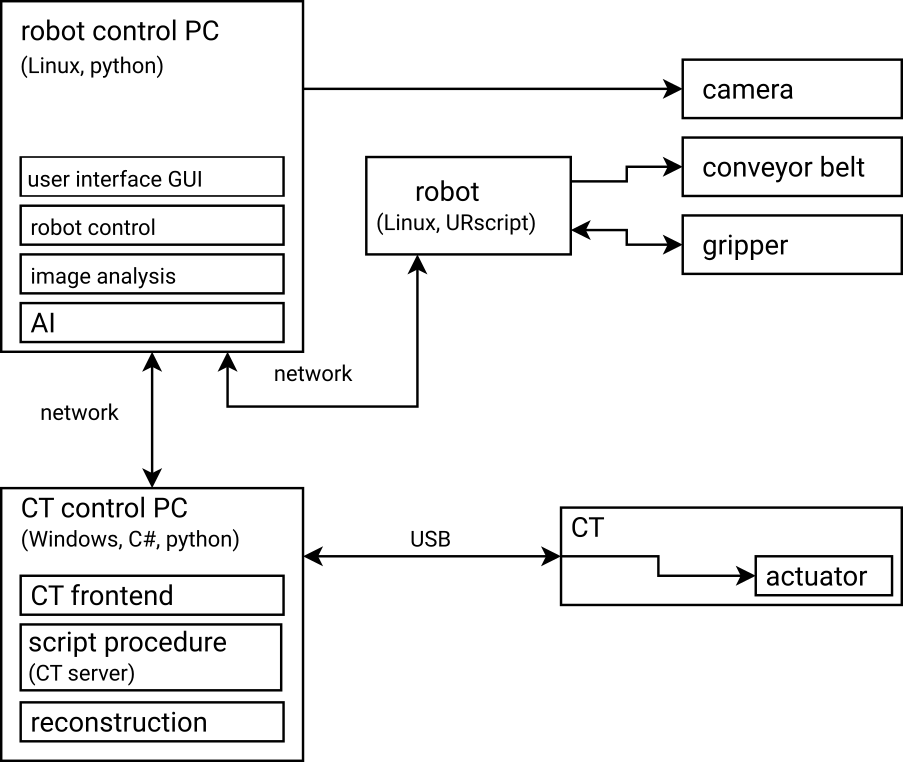}}
\caption{Flowchart of the analysis
}
\label{fig:schem}
\end{figure}

The apparatus comprises components for positioning and handling the vials, along with the tomograph itself. A schematic representation is provided in Fig. \ref{fig:cad}: The samples (A) are placed on a conveyor belt (B) and transported to a pickup area. A camera (D) positioned above the pickup area recognizes the presence of a sample, determines the vial's position, and reads its QR code. A robotic gripper (E, F) with a control interface (H) and controller (I) picks up the vial and deposits it inside the CT device (G), which has been equipped with an actuator to operate the safety cover. The radiograms of the vial captured by the tomograph are reconstructed into a full tomogram by the computer (C). The motion of the robotic arm and the automatic classification of the sample are managed by a second PC (D). To achieve high-quality tomograms, 400 individual radiograms are recorded for each sample, each boasting a resolution slightly exceeding $1$ million pixels and a color depth of 16 bits. Figure \ref{fig:schem} illustrates the connection of the process components.

\begin{figure}[ht]
\centerline{\includegraphics[width=\columnwidth]{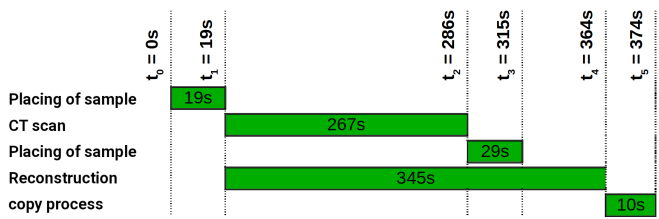}}
\caption{Flowchart depicting the timing of numerical processing.}
\label{fig:time}
\end{figure}

Although high throughput was not the primary objective of this conceptual study, we conducted a benchmark of the data acquisition process to identify potential areas for optimization. Figure \ref{fig:time} presents a flowchart depicting the time requirements for individual process steps. The evident bottlenecks are the CT-scan itself and the reconstruction of the tomogram.

In our setup, a single radiogram necessitates approximately 400\,ms of exposure time. This duration can be significantly reduced by employing a faster image sensor. Similarly, the reconstruction time can be diminished by utilizing more powerful CPU/GPU workstations. The same applies to the time required for copying and storing radiograms and performing volumetric reconstruction. Another option is to assess whether the number of radiograms recorded for each sample can be decreased.

It is worth noting that the automatic classification of the tomogram, as detailed in Section \ref{sec:automaticClassification}, only takes a few microseconds and is thus negligible in terms of the overall process time.

\section{\label{sec:automaticClassification}Automatic quality assessment}
\subsection{Concept}
We classify the samples based on their X-ray tomograms, as described in Sec. \ref{sec:hardware}. To achieve this, we export 400 equidistant horizontal slices (2D images) from the bottom to the top of the vial from the tomograms. Each 2D image undergoes classification for possible defects. From the resulting individual assessments, we calculate an overall rating for the sample.

For the classification of individual slices, we employ supervised machine learning. The objective is to identify a function capable of taking the image data of a 2D slice and providing the probability that the considered slice of the sample belongs to a specific class. In our case, examples of these classes could include \textit{major crack formation} or \textit{severe shrinkage}, as defined in Tab. \ref{tab:defect}. Technically, our input data consist of the grayscale values of each pixel in the 2D images, forming a matrix of grayscale values.

In the realm of machine learning, the functions used to predict labels from input data are commonly referred to as hypotheses. In supervised learning, hypotheses are typically determined by employing functions with numerous parameters. These parameters are chosen to predict the correct labels as accurately as possible for a given set of input data. The quality of this optimization is measured by a \emph{cost function}, quantifying the difference between the predicted labels and the ground truth labels. Various metrics, such as mean squared error or cross-entropy, can be utilized here.

In machine learning, the process of optimizing parameters is termed \emph{training}, and the set of labeled input data used is known as the \emph{training set}. We assume that the hypothesis derived from training can be generalized to data not included in the training set. To assess this generalization, another set of labeled input data, distinct from the training set, is required. This data is referred to as the \emph{test set}. The following two subsections describe the training set and the assumptions applied for classification.

\subsection{Training set}
The quality of any supervised learning algorithm heavily relies on the training set. A sufficiently large number of training examples is essential for each category to be classified, and ideally, these examples should be evenly distributed across different categories. Creating a comprehensive training set is challenging, especially when deliberately inducing defects during freeze-drying is difficult. Initially, we prepared a total of 720 samples from four different formulations of arginine, sucrose, and albumin-based systems at varying concentrations, as indicated in Tab. \ref{tab:formulations}. Formulations and drying conditions were selected based on prior experiences with these model systems \cite{startzel:2015a, startzel:2015b}, ensuring a wide range of expected defects.

The samples were prepared by measuring 6.5\,ml of the formulation into a standard 10R vial (dimensions OD $\times$ H: $24\,\text{mm}\times 45\,\text{mm}$). For freeze-drying, the vials were semi-sealed with a crimp neck stopper, and after the drying process, they were hermetically sealed. Subsequently, the finished samples underwent visual classification for defects outlined in Tab. \ref{tab:defect} by two operators. Both the actual sample and the CT sectional images, captured using the apparatus described in Sec. \ref{sec:hardware}, were examined. Results from individual inspections were compared, and samples with differing outcomes were re-inspected by both operators. During the preparation phase, each sample vial was marked with an individual QR code on its top for identification purposes.

\begin{table*}[ht]
\caption{Formulations of the training set\label{tab:formulations}}
\centering
\begin{tabular}{l|l|l|l|l|l|l|l}
formulation & arginine & saccharose & BSA &  histidine & polysorbate 80 & acid & pH \\
  &   [mg/mL]  &   [mg/mL]  &  [mg/mL] & [mg/mL]  &   [mg/mL] &    & \\
\hline
F1, 20\%  &  66.7  &  133.3 &   - &   3.1 &   0.2 & hydrochloric acid   & 6.0  \\
F1, 10\%  &  33.3  &  66.7  &  -  &  3.1  &  0.2  & hydrochloric acid     & 6.0 \\
F1, 5\%   & 16.7   & 33.3   & -   & 3.1   & 0.2   & hydrochloric acid      & 6.0  \\
F1, 1\%   & 3.3    & 6.7    & -   & 3.1   & 0.2   & hydrochloric acid      & 6.0  \\
F2, 20\%  &  66.7  & 133.3  &  -  &  3.1  & 0.2   & succinic acid & 6.0  \\
F2, 10\%  &  33.3  &  66.7  &  -  &  3.1  &  0.2  & succinic acid& 6.0  \\
F2, 5\%   & 16.7   & 33.3   & -   & 3.1   & 0.2   & succinic acid& 6.0  \\
F2, 1\%   & 3.3    & 6.7    & -   & 3.1   & 0.2   & succinic acid  &6.0  \\
F3, 20\%  & 180.0  & 20.0   & -   & 3.1   & 0.2   & hydrochloric acid      &6.0  \\
F3, 10\%  &  90.0  &  10.0  &  -  &  3.1  &  0.2  & hydrochloric acid     & 6.0  \\
F3, 5\%   & 45.0   & 5.0    & -   & 3.1   & 0.2   & hydrochloric acid     & 6.0  \\
F3, 1\%   & 9.0    & 1.0    & -   & 3.1   & 0.2   & hydrochloric acid      & 6.0  \\
F4, 10\%  &  61.5  &  -     & 38.5&    3.1&    0.2& hydrochloric acid  & 6.0  \\
F4, 5\%   & 30.8   & -      & 19.2&    3.1&    0.2& hydrochloric acid  & 6.0  \\
F4, 1\%   & 6.2    & -      & 3.8 &   3.1 &   0.2 & hydrochloric acid   & 6.0  \\
\end{tabular}
\end{table*}

The entire collection of visual classification outcomes is listed in the supplementary material. Here, we provide a summary of the key findings. It is evident that the concentration of the active ingredient appears to be the most crucial factor influencing both the frequency and variety of defects. The chemical composition of the formulation seems to play a subordinate role. The highest defect rate and variety are observed for a twenty percent concentration. The predominant defect scenario is shrinkage/collapse, followed by loose skin and foaming/bubble formation. Crack formation and blow-out occur at a much lower rate. It is important to note that the dataset exhibits an imbalance.
\begin{figure}[ht]
	\centerline{\includegraphics[width=\columnwidth]{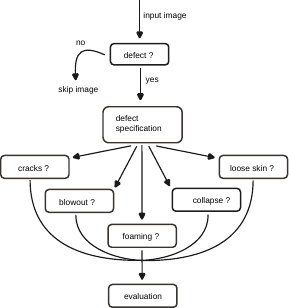}}
	\caption{Flowchart of the image analysis procedure\label{fig:aipipeline}}
\end{figure}

\subsection{Hypothesis/ learning algorithm}
The type of hypothesis that has proven highly effective for image classification is deep neural networks, especially convolutional neural networks (CNN). The architecture and functionality of these machine-learning algorithms are beyond the scope of this presentation. At this point, it suffices to understand that convolutional networks are complex functions that map multiple scalar input values to a single scalar output value, with the exact functional relationship influenced by freely selectable parameters. In the case of image classification, the input values include the grayscale values of individual pixels in the image, and the output value is a real number indicating the probability that the analyzed image belongs to a certain class.

To classify images with many thousands of pixels, CNNs contain a large number of parameters that need to be determined through training. In this work, we utilize the Inception v3 CNN architecture \cite{inception,Sam2019OfflineSVCGPT,Xia2017Inceptionv3FFCGPT,Wang2019PulmonaryICCGPT}, which comprises approximately 24 million parameters. Training a classifier with such a substantial number of parameters requires more training examples than our 720 samples provide. To address this challenge, we apply the concept of \emph{transfer learning} \cite{transfer,Lu2015TransferLUCGPT,Zhuang2020ComprehensiveSOCGPT}. Transfer learning leverages the fact that parameters learned for large parts of the network are relatively independent of specific image material, focusing on more general aspects of image processing, such as edge detection \cite{Gao2018DeepTLCGPT,Ghafoorian2017TransferLFCGPT,Minaee2020DeepCOVIDPCCGPT,faucris.258191566CGPT}. We start with the Inception v3 network pre-trained on the extensive ImageNet dataset \cite{imageNet,Russakovsky2015ImageNetLSCGPT,Kornblith2019DoBICGPT,Krizhevsky2017ImageNetCWCGPT}, consisting of more than 14 million labeled images. Our dataset is then used to train the end area of the network, which learns class-specific features. This process, known as \emph{re-training}, tailors the pre-trained Inception v3 framework to answer specific questions about the classification of freeze-dried products.

Our overall goal is to inspect the CT slice images of a sample vial for the defect categories described in Tab. \ref{tab:defect}. As shown in Fig. \ref{fig:aipipeline}, we distribute this task among six independently trained networks: The first network is trained to detect whether a defect occurs without specifying it precisely. If a defect is anticipated from this pre-analysis, the slice images of the vial are evaluated by five subsequent neural networks, each trained to detect the markedness of one of the defects from Tab. \ref{tab:defect}. This separation ensures a faster overall analysis and increases accuracy.

\begin{figure}[ht]
	\centerline{\includegraphics[width=0.95\columnwidth]{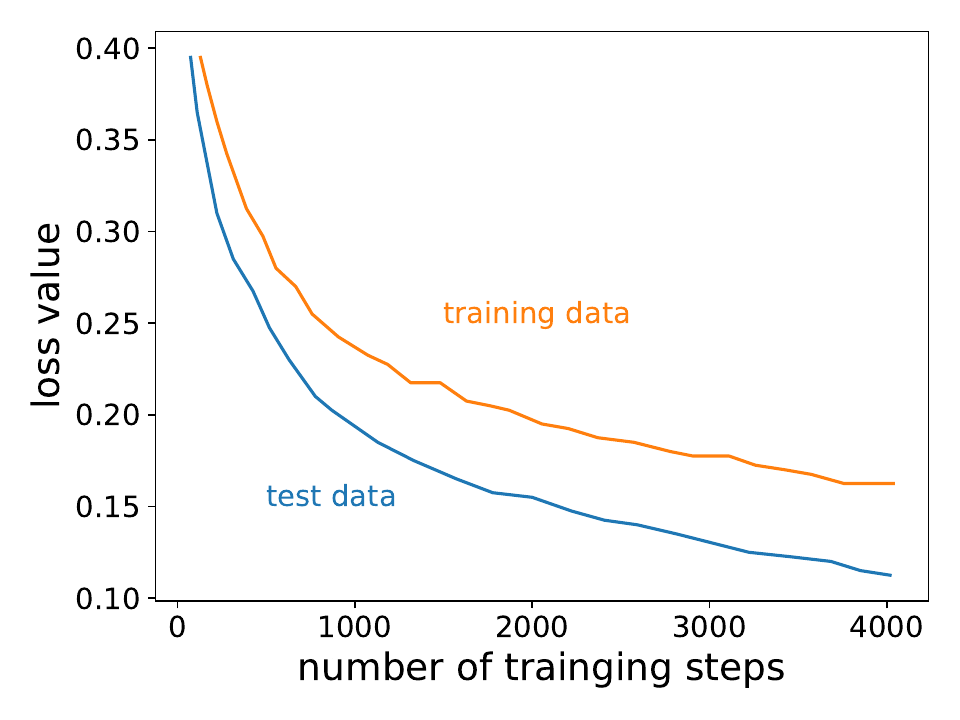}}
	\caption{Loss value (cost function) as a function of the number of training steps. Orange: loss for the training data; Blue: loss for the test data}
	\label{fig:loss}
\end{figure}

To monitor the training process, we examine the cost or loss function as a function of the training progress, i.e., the number of training steps. Here, we use the cross-entropy \cite{murphy:2012,Li2021MixedCECGPT,Ho2019TheRWCGPT} to calculate the loss. As seen in Fig. \ref{fig:loss}, the difference between the ground truth labels and the predicted labels decreases with the training progress for both the training-set data and the test-set data. This indicates a successful learning process, with learned parameters generalizing well to data not used for training.

\subsection{Performance of the learning algorithm}
To assess the performance of automatic quality assessment, we partition the available data into three segments: 80\% of the images serve as training data, 10\% of the images are used for hyperparameter tuning, such as the learning rate, and the remaining 10\% are employed as a final reality test. Due to the absence of certain error categories in samples with low drug concentrations, we limit this evaluation to data from samples with a 20\% active ingredient concentration, comprising 144 samples.

The performance of the learning algorithm varies significantly for two groups of defect categories: For the collapse, cracks, and foaming categories, we achieve a robust absolute accuracy of over 93\%, making it suitable for practical use. However, for the blowout and loose skin categories, the accuracy drops to approximately 50\%, indicating that automatic assessment does not function effectively. This can be attributed to three issues. It is important to note that, for this study, we exclusively consider vertical slices of the tomogram for classification. The defects of loose skin and, particularly, the blowout effect predominantly affect the vertical direction. This error could potentially be mitigated by additionally considering vertical slices. However, implementing this seemingly straightforward correction is challenging. Choosing slices away from the symmetry axis of the cylindrical sample geometry would cover different fractions of the vial depending on their distance from the symmetry axis. Conversely, selecting slices through the symmetry axis would result in an over-representation of data closer to the axis of the vial. This issue will be addressed in future enhancements of the machine learning algorithm.

In addition to the vertical nature of the defects (loose skin and blowout), which has not been considered thus far, these defects were under-represented in our dataset, further diminishing the accuracy of the learning algorithm. Finally, it must be acknowledged that these defects, especially their grading, are challenging to classify based on cross-sectional CT images, even for an experienced individual.

\section{Conclusion and outlook}

We present a proof of concept demonstrating the feasibility of fully automated assessment of the structural quality of freeze-dried pharmaceuticals using X-ray tomography. We have outlined an apparatus capable of feeding, scanning, and separating freeze-dried samples based on their assessments. Additionally, we have introduced a machine-learning algorithm for automatically classifying samples according to their X-ray tomograms. Once the learning algorithm is trained, the system can operate with minimal human interaction.

It is important to note that the described setup is not immediately suitable for industrial-scale applications, where even small product batches consist of several thousand vials. However, we have established that the fundamental concept of the entire process chain is viable and has the potential for future scaling to an industrial level. Potential avenues for achieving this include reducing the number of X-ray projections used for each tomogram, employing more sensitive detectors to decrease the exposure time for each radiogram, and exploring the possibilities of parallelizing the process or implementing continuous process tomographic methods.

In this study, the machine learning algorithm encountered challenges in reliably classifying certain defects listed in Table \ref{tab:defect}, which are pertinent for regulatory approval. However, this limitation can be ascribed to the inadequacy of the training dataset and does not represent a fundamental flaw in our concept. The current availability of an automatic setup allows for the generation of much more extensive data records. In future iterations, we intend to incorporate vertical CT-slices of the sample, a modification expected to notably improve the accuracy of classification for the defect categories \textit{blowout} and \textit{loose skin}.

The approval documents for freeze-dried products typically specify a specific visual appearance. In this study, we focused on analyzing the product structure. However, other attributes related to product quality, especially the reconstitution behavior of the lyophilisate, can be more crucial in practical applications than the product's appearance. Ultimately, the decisive factor is the quality of the reconstituted lyophilisate rather than its visual characteristics.

A promising avenue for future research is to develop a learning algorithm capable of predicting the reconstitution behavior of freeze-dried products based on their computed tomography images. In this approach, the slices of the tomograms in the corresponding training set should be labeled according to attributes such as reconstitution time or the quality of the reconstituted product, rather than structural properties like the presence of cracks in the lyophilisate.

\section{Acknowledgments}
We acknowledge the funding received from the German Science Foundation (DFG) through the Interdisciplinary Centre for Nanostructured Films (IZNF). Our gratitude goes to Walter Pucheanu for his technical support. ChatGPT \cite{ChatGPT} was used to improve English writing.

\bibliography{medac.bib}

\begin{thebibliography}{53}%
\makeatletter
\providecommand \@ifxundefined [1]{%
 \@ifx{#1\undefined}
}%
\providecommand \@ifnum [1]{%
 \ifnum #1\expandafter \@firstoftwo
 \else \expandafter \@secondoftwo
 \fi
}%
\providecommand \@ifx [1]{%
 \ifx #1\expandafter \@firstoftwo
 \else \expandafter \@secondoftwo
 \fi
}%
\providecommand \natexlab [1]{#1}%
\providecommand \enquote  [1]{``#1''}%
\providecommand \bibnamefont  [1]{#1}%
\providecommand \bibfnamefont [1]{#1}%
\providecommand \citenamefont [1]{#1}%
\providecommand \href@noop [0]{\@secondoftwo}%
\providecommand \href [0]{\begingroup \@sanitize@url \@href}%
\providecommand \@href[1]{\@@startlink{#1}\@@href}%
\providecommand \@@href[1]{\endgroup#1\@@endlink}%
\providecommand \@sanitize@url [0]{\catcode `\\12\catcode `\$12\catcode
  `\&12\catcode `\#12\catcode `\^12\catcode `\_12\catcode `\%12\relax}%
\providecommand \@@startlink[1]{}%
\providecommand \@@endlink[0]{}%
\providecommand \url  [0]{\begingroup\@sanitize@url \@url }%
\providecommand \@url [1]{\endgroup\@href {#1}{\urlprefix }}%
\providecommand \urlprefix  [0]{URL }%
\providecommand \Eprint [0]{\href }%
\providecommand \doibase [0]{http://dx.doi.org/}%
\providecommand \selectlanguage [0]{\@gobble}%
\providecommand \bibinfo  [0]{\@secondoftwo}%
\providecommand \bibfield  [0]{\@secondoftwo}%
\providecommand \translation [1]{[#1]}%
\providecommand \BibitemOpen [0]{}%
\providecommand \bibitemStop [0]{}%
\providecommand \bibitemNoStop [0]{.\EOS\space}%
\providecommand \EOS [0]{\spacefactor3000\relax}%
\providecommand \BibitemShut  [1]{\csname bibitem#1\endcsname}%
\let\auto@bib@innerbib\@empty
\bibitem [{\citenamefont {Pikal}(2010)}]{pikal:2010}%
  \BibitemOpen
  \bibfield  {author} {\bibinfo {author} {\bibfnamefont {J.~M.}\ \bibnamefont
  {Pikal}},\ }\enquote {\bibinfo {title} {Mechanisms of protein stabilization
  during freeze-drying storage: the relative importance of thermodynamic
  stabilization and glassy state relaxation dynamics},}\ in\ \href {\doibase
  10.3109/9781439825761} {\emph {\bibinfo {booktitle}
  {Freeze-Drying/Lyophilization of Pharmaceutical \& Biological Products}}}\
  (\bibinfo  {publisher} {CRC Press},\ \bibinfo {year} {2010})\ pp.\ \bibinfo
  {pages} {161--198},\ \bibinfo {edition} {3rd}\ ed.\BibitemShut {Stop}%
\bibitem [{\citenamefont {Wang}(2000)}]{wang:2000}%
  \BibitemOpen
  \bibfield  {author} {\bibinfo {author} {\bibfnamefont {W.}~\bibnamefont
  {Wang}},\ }\bibfield  {title} {\enquote {\bibinfo {title} {Lyophilization and
  development of solid protein pharmaceuticals},}\ }\href {\doibase
  10.1016/S0378-5173(00)00423-3} {\bibfield  {journal} {\bibinfo  {journal}
  {International Journal of Pharmaceutics}\ }\textbf {\bibinfo {volume}
  {203}},\ \bibinfo {pages} {1--60} (\bibinfo {year} {2000})}\BibitemShut
  {NoStop}%
\bibitem [{\citenamefont {Härdter}\ \emph {et~al.}(2023)\citenamefont
  {Härdter}, \citenamefont {Geidobler}, \citenamefont {Presser},\ and\
  \citenamefont {Winter}}]{Hardter2023CGPT}%
  \BibitemOpen
  \bibfield  {author} {\bibinfo {author} {\bibfnamefont {N.}~\bibnamefont
  {Härdter}}, \bibinfo {author} {\bibfnamefont {R.}~\bibnamefont {Geidobler}},
  \bibinfo {author} {\bibfnamefont {I.}~\bibnamefont {Presser}}, \ and\
  \bibinfo {author} {\bibfnamefont {G.}~\bibnamefont {Winter}},\ }\bibfield
  {title} {\enquote {\bibinfo {title} {Microwave-assisted freeze–drying:
  {I}mpact of microwave radiation on the quality of high-concentration antibody
  formulations},}\ }\href {\doibase 10.3390/pharmaceutics15122783} {\bibfield
  {journal} {\bibinfo  {journal} {Pharmaceutics}\ }\textbf {\bibinfo {volume}
  {15}},\ \bibinfo {pages} {2783} (\bibinfo {year} {2023})}\BibitemShut
  {NoStop}%
\bibitem [{\citenamefont {Pardeshi}\ \emph {et~al.}(2023)\citenamefont
  {Pardeshi}, \citenamefont {Deshmukh}, \citenamefont {Telange}, \citenamefont
  {Nangare}, \citenamefont {Sonar}, \citenamefont {Lakade}, \citenamefont
  {Harde}, \citenamefont {Pardeshi}, \citenamefont {Gholap}, \citenamefont
  {Deshmukh},\ and\ \citenamefont {More}}]{Pardeshi2023CGPT}%
  \BibitemOpen
  \bibfield  {author} {\bibinfo {author} {\bibfnamefont {S.~R.}\ \bibnamefont
  {Pardeshi}}, \bibinfo {author} {\bibfnamefont {N.~S.}\ \bibnamefont
  {Deshmukh}}, \bibinfo {author} {\bibfnamefont {D.~R.}\ \bibnamefont
  {Telange}}, \bibinfo {author} {\bibfnamefont {Sopan~N.}\ \bibnamefont
  {Nangare}}, \bibinfo {author} {\bibfnamefont {Yogesh~Y.}\ \bibnamefont
  {Sonar}}, \bibinfo {author} {\bibfnamefont {Sameer~H.}\ \bibnamefont
  {Lakade}}, \bibinfo {author} {\bibfnamefont {Minal~T.}\ \bibnamefont
  {Harde}}, \bibinfo {author} {\bibfnamefont {Chandrakantsing~V.}\ \bibnamefont
  {Pardeshi}}, \bibinfo {author} {\bibfnamefont {Amol}\ \bibnamefont {Gholap}},
  \bibinfo {author} {\bibfnamefont {Prashant~K.}\ \bibnamefont {Deshmukh}}, \
  and\ \bibinfo {author} {\bibfnamefont {Mahesh~P.}\ \bibnamefont {More}},\
  }\bibfield  {title} {\enquote {\bibinfo {title} {Process development and
  quality attributes for the freeze-drying process in pharmaceuticals,
  biopharmaceuticals and nanomedicine delivery: a state-of-the-art review},}\
  }\href {\doibase 10.1186/s43094-023-00551-8} {\bibfield  {journal} {\bibinfo
  {journal} {Journal of Pharmaceutical Innovation}\ }\textbf {\bibinfo {volume}
  {2023}} (\bibinfo {year} {2023}),\ 10.1186/s43094-023-00551-8}\BibitemShut
  {NoStop}%
\bibitem [{\citenamefont {Manning}\ \emph {et~al.}(2010)\citenamefont
  {Manning}, \citenamefont {Chou}, \citenamefont {Murphy}, \citenamefont
  {Payne},\ and\ \citenamefont {Katayama}}]{Manning:2010}%
  \BibitemOpen
  \bibfield  {author} {\bibinfo {author} {\bibfnamefont {M.~C.}\ \bibnamefont
  {Manning}}, \bibinfo {author} {\bibfnamefont {D.~K.}\ \bibnamefont {Chou}},
  \bibinfo {author} {\bibfnamefont {B.~M.}\ \bibnamefont {Murphy}}, \bibinfo
  {author} {\bibfnamefont {R.~W.}\ \bibnamefont {Payne}}, \ and\ \bibinfo
  {author} {\bibfnamefont {D.~S.}\ \bibnamefont {Katayama}},\ }\bibfield
  {title} {\enquote {\bibinfo {title} {Stability of protein pharmaceuticals: an
  update.}}\ }\href {\doibase 10.1007/s11095-009-0045-6} {\bibfield  {journal}
  {\bibinfo  {journal} {Pharm Res}\ }\textbf {\bibinfo {volume} {27}},\
  \bibinfo {pages} {544--575} (\bibinfo {year} {2010})}\BibitemShut {NoStop}%
\bibitem [{\citenamefont {Rebizzi}(2023)}]{Rebizzi2023CGPT}%
  \BibitemOpen
  \bibfield  {author} {\bibinfo {author} {\bibfnamefont {M.}~\bibnamefont
  {Rebizzi}},\ }\emph {\bibinfo {title} {Development of releasable PEG for
  transient protein PEGylation}},\ \href
  {https://thesis.unipd.it/handle/20.500.12608/62182} {Ph.D. thesis},\ \bibinfo
   {school} {University of Padova} (\bibinfo {year} {2023})\BibitemShut
  {NoStop}%
\bibitem [{\citenamefont {Carpenter}\ and\ \citenamefont
  {Chang}(2020)}]{Carpenter2020LyophilizationOPCGPT}%
  \BibitemOpen
  \bibfield  {author} {\bibinfo {author} {\bibfnamefont {J.~F.}\ \bibnamefont
  {Carpenter}}\ and\ \bibinfo {author} {\bibfnamefont {B.~S.}\ \bibnamefont
  {Chang}},\ }\enquote {\bibinfo {title} {Lyophilization of protein
  pharmaceuticals},}\ in\ \href@noop {} {\emph {\bibinfo {booktitle}
  {Biotechnology and Biopharmaceutical Manufacturing, Processing, and
  Preservation}}}\ (\bibinfo  {publisher} {Taylor \& Francis},\ \bibinfo {year}
  {2020})\BibitemShut {NoStop}%
\bibitem [{\citenamefont {Remmele}\ \emph {et~al.}(2012)\citenamefont
  {Remmele}, \citenamefont {Krishnan}, \citenamefont {Ritter}, \citenamefont
  {Callahan},\ and\ \citenamefont {Warne}}]{Remmele2012DevelopmentOSCGPT}%
  \BibitemOpen
  \bibfield  {author} {\bibinfo {author} {\bibfnamefont {R.~L.}\ \bibnamefont
  {Remmele}}, \bibinfo {author} {\bibfnamefont {S.}~\bibnamefont {Krishnan}},
  \bibinfo {author} {\bibfnamefont {N.~M.}\ \bibnamefont {Ritter}}, \bibinfo
  {author} {\bibfnamefont {W.~L.}\ \bibnamefont {Callahan}}, \ and\ \bibinfo
  {author} {\bibfnamefont {J.~M.}\ \bibnamefont {Warne}},\ }\bibfield  {title}
  {\enquote {\bibinfo {title} {Development of stable lyophilized protein drug
  products},}\ }\href {\doibase 10.2174/138920112799361937} {\bibfield
  {journal} {\bibinfo  {journal} {Current Pharmaceutical Biotechnology}\
  }\textbf {\bibinfo {volume} {13}},\ \bibinfo {pages} {471--496} (\bibinfo
  {year} {2012})}\BibitemShut {NoStop}%
\bibitem [{\citenamefont {Pikal}(2007)}]{pikal:2007}%
  \BibitemOpen
  \bibfield  {author} {\bibinfo {author} {\bibfnamefont {M.~J.}\ \bibnamefont
  {Pikal}},\ }\enquote {\bibinfo {title} {Freeze drying},}\ in\ \href@noop {}
  {\emph {\bibinfo {booktitle} {Encyclopedia of Pharmaceutical Technology}}},\
  Vol.~\bibinfo {volume} {3},\ \bibinfo {editor} {edited by\ \bibinfo {editor}
  {\bibfnamefont {J.}~\bibnamefont {Swarbrick}}}\ (\bibinfo  {publisher}
  {Informa Healthcare},\ \bibinfo {address} {London},\ \bibinfo {year} {2007})\
  pp.\ \bibinfo {pages} {1807--1833},\ \bibinfo {edition} {3rd}\
  ed.\BibitemShut {Stop}%
\bibitem [{\citenamefont {Tang}\ and\ \citenamefont {Pikal}(2004)}]{tang:2004}%
  \BibitemOpen
  \bibfield  {author} {\bibinfo {author} {\bibfnamefont {X.}~\bibnamefont
  {Tang}}\ and\ \bibinfo {author} {\bibfnamefont {M.}~\bibnamefont {Pikal}},\
  }\bibfield  {title} {\enquote {\bibinfo {title} {Design of freeze-drying
  processes for pharmaceuticals: {P}ractical advice},}\ }\href {\doibase
  10.1023/B:PHAM.0000016234.73023.75} {\bibfield  {journal} {\bibinfo
  {journal} {Pharmaceutical Research}\ }\textbf {\bibinfo {volume} {21}},\
  \bibinfo {pages} {191--200} (\bibinfo {year} {2004})}\BibitemShut {NoStop}%
\bibitem [{\citenamefont {Liu}\ \emph {et~al.}(2005)\citenamefont {Liu},
  \citenamefont {Viverette}, \citenamefont {Virgin}, \citenamefont {Anderson},\
  and\ \citenamefont {Dalal}}]{Liu2005StudyOTCGPT}%
  \BibitemOpen
  \bibfield  {author} {\bibinfo {author} {\bibfnamefont {J.}~\bibnamefont
  {Liu}}, \bibinfo {author} {\bibfnamefont {T.}~\bibnamefont {Viverette}},
  \bibinfo {author} {\bibfnamefont {M.}~\bibnamefont {Virgin}}, \bibinfo
  {author} {\bibfnamefont {M.}~\bibnamefont {Anderson}}, \ and\ \bibinfo
  {author} {\bibfnamefont {P.}~\bibnamefont {Dalal}},\ }\bibfield  {title}
  {\enquote {\bibinfo {title} {A study of the impact of freezing on the
  lyophilization of a concentrated formulation with a high fill depth},}\
  }\href {\doibase 10.1081/PDT-54452} {\bibfield  {journal} {\bibinfo
  {journal} {Pharmaceutical Development and Technology}\ }\textbf {\bibinfo
  {volume} {10}},\ \bibinfo {pages} {261--272} (\bibinfo {year}
  {2005})}\BibitemShut {NoStop}%
\bibitem [{\citenamefont {Butreddy}\ \emph {et~al.}(2020)\citenamefont
  {Butreddy}, \citenamefont {Dudhipala}, \citenamefont {Janga},\ and\
  \citenamefont {Gaddam}}]{Butreddy2020LyophilizationOSCGPT}%
  \BibitemOpen
  \bibfield  {author} {\bibinfo {author} {\bibfnamefont {A.}~\bibnamefont
  {Butreddy}}, \bibinfo {author} {\bibfnamefont {N.}~\bibnamefont {Dudhipala}},
  \bibinfo {author} {\bibfnamefont {K.~Y.}\ \bibnamefont {Janga}}, \ and\
  \bibinfo {author} {\bibfnamefont {R.~P.}\ \bibnamefont {Gaddam}},\ }\bibfield
   {title} {\enquote {\bibinfo {title} {Lyophilization of small-molecule
  injectables: {A}n industry perspective on formulation development, process
  optimization, scale-up challenges, and drug product quality attributes},}\
  }\href {\doibase 10.1208/s12249-020-01787-w} {\bibfield  {journal} {\bibinfo
  {journal} {AAPS PharmSciTech}\ }\textbf {\bibinfo {volume} {21}},\ \bibinfo
  {pages} {222} (\bibinfo {year} {2020})}\BibitemShut {NoStop}%
\bibitem [{\citenamefont {Barresi}\ \emph {et~al.}(2009)\citenamefont
  {Barresi}, \citenamefont {Ghio}, \citenamefont {Fissore},\ and\ \citenamefont
  {Pisano}}]{barresi:2009}%
  \BibitemOpen
  \bibfield  {author} {\bibinfo {author} {\bibfnamefont {A.A.}\ \bibnamefont
  {Barresi}}, \bibinfo {author} {\bibfnamefont {S.}~\bibnamefont {Ghio}},
  \bibinfo {author} {\bibfnamefont {D.}~\bibnamefont {Fissore}}, \ and\
  \bibinfo {author} {\bibfnamefont {R.}~\bibnamefont {Pisano}},\ }\bibfield
  {title} {\enquote {\bibinfo {title} {Freeze drying of pharmaceutical
  excipients close to collapse temperature: {I}nfluence of the process
  conditions on process time and product quality},}\ }\href {\doibase
  10.1080/07373930902901646} {\bibfield  {journal} {\bibinfo  {journal} {Dry.
  Technol.}\ }\textbf {\bibinfo {volume} {27}},\ \bibinfo {pages} {805--16}
  (\bibinfo {year} {2009})}\BibitemShut {NoStop}%
\bibitem [{ins(2014)}]{inspection:2014}%
  \BibitemOpen
  \href
  {https://www.fda.gov/inspections-compliance-enforcement-and-criminal-investigations/inspection-guides/lyophilization-parenteral-793}
  {\emph {\bibinfo {title} {Inspection Technical Guides: {L}yophilization of
  Parenteral}}},\ \bibinfo {address} {USA} (\bibinfo {year} {2014})\BibitemShut
  {NoStop}%
\bibitem [{\citenamefont {Patel}\ \emph {et~al.}(2017)\citenamefont {Patel},
  \citenamefont {Nail}, \citenamefont {Pikal}, \citenamefont {Geidobler},
  \citenamefont {Winter}, \citenamefont {Hawe}, \citenamefont {Davagnino},\
  and\ \citenamefont {Gupta}}]{patel:2017}%
  \BibitemOpen
  \bibfield  {author} {\bibinfo {author} {\bibfnamefont {S.~M.}\ \bibnamefont
  {Patel}}, \bibinfo {author} {\bibfnamefont {S.~L.}\ \bibnamefont {Nail}},
  \bibinfo {author} {\bibfnamefont {M.~J.}\ \bibnamefont {Pikal}}, \bibinfo
  {author} {\bibfnamefont {R.}~\bibnamefont {Geidobler}}, \bibinfo {author}
  {\bibfnamefont {G.}~\bibnamefont {Winter}}, \bibinfo {author} {\bibfnamefont
  {A.}~\bibnamefont {Hawe}}, \bibinfo {author} {\bibfnamefont {J.}~\bibnamefont
  {Davagnino}}, \ and\ \bibinfo {author} {\bibfnamefont {S.~R.}\ \bibnamefont
  {Gupta}},\ }\bibfield  {title} {\enquote {\bibinfo {title} {Lyophilized drug
  product cake appearance: What is acceptable?}}\ }\href {\doibase
  10.1016/j.xphs.2017.03.014} {\bibfield  {journal} {\bibinfo  {journal} {J.
  Pharm. Sci.}\ }\textbf {\bibinfo {volume} {106}},\ \bibinfo {pages}
  {1706--1721} (\bibinfo {year} {2017})}\BibitemShut {NoStop}%
\bibitem [{fda(2014)}]{fda:2014}%
  \BibitemOpen
  \href
  {https://www.fda.gov/inspections-compliance-enforcement-and-criminal-investigations/inspection-guides/lyophilization-parenteral-793}
  {\enquote {\bibinfo {title} {Lyophilization of parenteral (7/93) - guide to
  inspections of lyophilization of parenterals},}\ }\bibinfo {howpublished}
  {https://www.fda.gov/inspections-compliance-enforcement-and-criminal-investigations/inspection-guides/lyophilization-parenteral-793}
  (\bibinfo {year} {2014})\BibitemShut {NoStop}%
\bibitem [{\citenamefont {Rambhatla}\ \emph {et~al.}(2004)\citenamefont
  {Rambhatla}, \citenamefont {Ramot}, \citenamefont {Bhugra},\ and\
  \citenamefont {Pikal}}]{rambhatla:2004}%
  \BibitemOpen
  \bibfield  {author} {\bibinfo {author} {\bibfnamefont {S.}~\bibnamefont
  {Rambhatla}}, \bibinfo {author} {\bibfnamefont {R.}~\bibnamefont {Ramot}},
  \bibinfo {author} {\bibfnamefont {C.}~\bibnamefont {Bhugra}}, \ and\ \bibinfo
  {author} {\bibfnamefont {M.~J.}\ \bibnamefont {Pikal}},\ }\bibfield  {title}
  {\enquote {\bibinfo {title} {Heat and mass transfer scale-up issues during
  freeze drying: {II. C}ontrol and characterization of the degree of
  supercooling},}\ }\href {\doibase 10.1208/pt050458} {\bibfield  {journal}
  {\bibinfo  {journal} {AAPS PharmSciTech}\ }\textbf {\bibinfo {volume} {5}},\
  \bibinfo {pages} {54–62} (\bibinfo {year} {2004})}\BibitemShut {NoStop}%
\bibitem [{\citenamefont {Schersch}\ \emph {et~al.}(2010)\citenamefont
  {Schersch}, \citenamefont {Betz}, \citenamefont {Garidel}, \citenamefont
  {Muehlau}, \citenamefont {Bassarab},\ and\ \citenamefont
  {Winter}}]{schersch:2010}%
  \BibitemOpen
  \bibfield  {author} {\bibinfo {author} {\bibfnamefont {K.}~\bibnamefont
  {Schersch}}, \bibinfo {author} {\bibfnamefont {O.}~\bibnamefont {Betz}},
  \bibinfo {author} {\bibfnamefont {P.}~\bibnamefont {Garidel}}, \bibinfo
  {author} {\bibfnamefont {S.}~\bibnamefont {Muehlau}}, \bibinfo {author}
  {\bibfnamefont {S.}~\bibnamefont {Bassarab}}, \ and\ \bibinfo {author}
  {\bibfnamefont {G.}~\bibnamefont {Winter}},\ }\bibfield  {title} {\enquote
  {\bibinfo {title} {Systematic investigation of the effect of lyophilizate
  collapse on pharmaceutically relevant proteins i: Stability after
  freeze‐drying},}\ }\href {\doibase 10.1002/jps.22000} {\bibfield  {journal}
  {\bibinfo  {journal} {Journal of Pharmaceutical Sciences}\ }\textbf {\bibinfo
  {volume} {99}},\ \bibinfo {pages} {2256--2278} (\bibinfo {year}
  {2010})}\BibitemShut {NoStop}%
\bibitem [{\citenamefont {Liu}(2006)}]{liu:2006}%
  \BibitemOpen
  \bibfield  {author} {\bibinfo {author} {\bibfnamefont {J.}~\bibnamefont
  {Liu}},\ }\bibfield  {title} {\enquote {\bibinfo {title} {Physical
  characterization of pharmaceutical formulations in frozen and freeze-dried
  solid states: {T}echniques and applications in freeze-drying development},}\
  }\href {\doibase 10.1080/10837450500463729} {\bibfield  {journal} {\bibinfo
  {journal} {Pharmaceutical Development and Technology}\ }\textbf {\bibinfo
  {volume} {11}},\ \bibinfo {pages} {3--28} (\bibinfo {year} {2006})},\ \Eprint
  {http://arxiv.org/abs/https://doi.org/10.1080/10837450500463729}
  {https://doi.org/10.1080/10837450500463729} \BibitemShut {NoStop}%
\bibitem [{\citenamefont {Kim}\ \emph {et~al.}(1998)\citenamefont {Kim},
  \citenamefont {Akers},\ and\ \citenamefont {Nail}}]{kim:1998}%
  \BibitemOpen
  \bibfield  {author} {\bibinfo {author} {\bibfnamefont {Alexandra~I.}\
  \bibnamefont {Kim}}, \bibinfo {author} {\bibfnamefont {Michael~J.}\
  \bibnamefont {Akers}}, \ and\ \bibinfo {author} {\bibfnamefont {Steven~L.}\
  \bibnamefont {Nail}},\ }\bibfield  {title} {\enquote {\bibinfo {title} {The
  physical state of mannitol after freeze-drying: {E}ffects of mannitol
  concentration, freezing rate, and a noncrystallizing cosolute},}\ }\href
  {\doibase 10.1021/js980001d} {\bibfield  {journal} {\bibinfo  {journal}
  {Journal of Pharmaceutical Sciences}\ }\textbf {\bibinfo {volume} {87}},\
  \bibinfo {pages} {931--935} (\bibinfo {year} {1998})}\BibitemShut {NoStop}%
\bibitem [{\citenamefont {Stärtzel}\ \emph
  {et~al.}(2015{\natexlab{a}})\citenamefont {Stärtzel}, \citenamefont
  {Gieseler}, \citenamefont {Gieseler}, \citenamefont {Abdul-Fattah},
  \citenamefont {Adler}, \citenamefont {Mahler},\ and\ \citenamefont
  {Goldbach}}]{startzel:2015b}%
  \BibitemOpen
  \bibfield  {author} {\bibinfo {author} {\bibfnamefont {P.}~\bibnamefont
  {Stärtzel}}, \bibinfo {author} {\bibfnamefont {H.}~\bibnamefont {Gieseler}},
  \bibinfo {author} {\bibfnamefont {M.}~\bibnamefont {Gieseler}}, \bibinfo
  {author} {\bibfnamefont {A.M.}\ \bibnamefont {Abdul-Fattah}}, \bibinfo
  {author} {\bibfnamefont {M.}~\bibnamefont {Adler}}, \bibinfo {author}
  {\bibfnamefont {H.C.}\ \bibnamefont {Mahler}}, \ and\ \bibinfo {author}
  {\bibfnamefont {P.}~\bibnamefont {Goldbach}},\ }\bibfield  {title} {\enquote
  {\bibinfo {title} {Freeze drying of l-arginine/sucrose-based protein
  formulations, part 2: Optimization of formulation design and freeze-drying
  process conditions for an l-arginine chloride-based protein formulation
  system},}\ }\href {\doibase 10.1002/jps.24658} {\bibfield  {journal}
  {\bibinfo  {journal} {J. Pharm. Sci.}\ }\textbf {\bibinfo {volume} {104}},\
  \bibinfo {pages} {4241--56} (\bibinfo {year}
  {2015}{\natexlab{a}})}\BibitemShut {NoStop}%
\bibitem [{\citenamefont {Geidobler}\ \emph {et~al.}(2013)\citenamefont
  {Geidobler}, \citenamefont {Konrad},\ and\ \citenamefont
  {Winter}}]{geidobler:2013}%
  \BibitemOpen
  \bibfield  {author} {\bibinfo {author} {\bibfnamefont {R.}~\bibnamefont
  {Geidobler}}, \bibinfo {author} {\bibfnamefont {I.}~\bibnamefont {Konrad}}, \
  and\ \bibinfo {author} {\bibfnamefont {G.}~\bibnamefont {Winter}},\
  }\bibfield  {title} {\enquote {\bibinfo {title} {Can controlled ice
  nucleation improve freeze‐drying of highly‐concentrated protein
  formulations?}}\ }\href {\doibase 10.1002/jps.23704} {\bibfield  {journal}
  {\bibinfo  {journal} {Journal of Pharmaceutical Sciences}\ }\textbf {\bibinfo
  {volume} {102}},\ \bibinfo {pages} {3915--3919} (\bibinfo {year}
  {2013})}\BibitemShut {NoStop}%
\bibitem [{\citenamefont {Tsay}\ and\ \citenamefont {Li}(2019)}]{tsay:2019}%
  \BibitemOpen
  \bibfield  {author} {\bibinfo {author} {\bibfnamefont {C.}~\bibnamefont
  {Tsay}}\ and\ \bibinfo {author} {\bibfnamefont {Z.}~\bibnamefont {Li}},\
  }\bibfield  {title} {\enquote {\bibinfo {title} {Automating visual inspection
  of lyophilized drug products with multi-input deep neural networks},}\ }in\
  \href@noop {} {\emph {\bibinfo {booktitle} {2019 IEEE 15th International
  Conference on Automation Science and Engineering (CASE)}}}\ (\bibinfo
  {organization} {IEEE},\ \bibinfo {year} {2019})\ pp.\ \bibinfo {pages}
  {1802--1807}\BibitemShut {NoStop}%
\bibitem [{\citenamefont {Haeuser}\ \emph {et~al.}(2018)\citenamefont
  {Haeuser}, \citenamefont {Goldbach}, \citenamefont {Huwyler}, \citenamefont
  {Friess},\ and\ \citenamefont {Allmendinger}}]{haeuser:2018}%
  \BibitemOpen
  \bibfield  {author} {\bibinfo {author} {\bibfnamefont {C.}~\bibnamefont
  {Haeuser}}, \bibinfo {author} {\bibfnamefont {P.}~\bibnamefont {Goldbach}},
  \bibinfo {author} {\bibfnamefont {J.}~\bibnamefont {Huwyler}}, \bibinfo
  {author} {\bibfnamefont {W.}~\bibnamefont {Friess}}, \ and\ \bibinfo {author}
  {\bibfnamefont {A.}~\bibnamefont {Allmendinger}},\ }\bibfield  {title}
  {\enquote {\bibinfo {title} {Imaging techniques to characterize cake
  appearance of freeze-dried products},}\ }\href {\doibase
  10.1016/j.xphs.2018.06.025} {\bibfield  {journal} {\bibinfo  {journal}
  {Journal of Pharmaceutical Sciences}\ }\textbf {\bibinfo {volume} {107}},\
  \bibinfo {pages} {2810--2822} (\bibinfo {year} {2018})}\BibitemShut {NoStop}%
\bibitem [{\citenamefont {R.}\ \emph {et~al.}(2017)\citenamefont {R.},
  \citenamefont {Barresi}, \citenamefont {Capozzi}, \citenamefont {Novajra},
  \citenamefont {Oddone},\ and\ \citenamefont
  {Vitale-Brovarone}}]{pisano:2017}%
  \BibitemOpen
  \bibfield  {author} {\bibinfo {author} {\bibfnamefont {Pisano}\ \bibnamefont
  {R.}}, \bibinfo {author} {\bibfnamefont {A.A.}\ \bibnamefont {Barresi}},
  \bibinfo {author} {\bibfnamefont {L.}~\bibnamefont {Capozzi}}, \bibinfo
  {author} {\bibfnamefont {G.}~\bibnamefont {Novajra}}, \bibinfo {author}
  {\bibfnamefont {I.}~\bibnamefont {Oddone}}, \ and\ \bibinfo {author}
  {\bibfnamefont {C.}~\bibnamefont {Vitale-Brovarone}},\ }\bibfield  {title}
  {\enquote {\bibinfo {title} {Characterization of the mass transfer of
  lyophilized products based on x-ray micro-computed tomography images},}\
  }\href {\doibase doi.org/10.1080/07373937.2016.1222540} {\bibfield  {journal}
  {\bibinfo  {journal} {Dry. Technol.}\ }\textbf {\bibinfo {volume} {35}},\
  \bibinfo {pages} {933--938} (\bibinfo {year} {2017})}\BibitemShut {NoStop}%
\bibitem [{\citenamefont {Kunz}\ \emph {et~al.}(2019)\citenamefont {Kunz},
  \citenamefont {Schuldt-Lieb},\ and\ \citenamefont {Gieseler}}]{kunz:2019}%
  \BibitemOpen
  \bibfield  {author} {\bibinfo {author} {\bibfnamefont {C.}~\bibnamefont
  {Kunz}}, \bibinfo {author} {\bibfnamefont {S.}~\bibnamefont {Schuldt-Lieb}},
  \ and\ \bibinfo {author} {\bibfnamefont {H.}~\bibnamefont {Gieseler}},\
  }\bibfield  {title} {\enquote {\bibinfo {title} {Freeze-drying from organic
  co-solvent systems, part 2: Process modifications to reduce residual solvent
  levels and improve product quality attributes},}\ }\href {\doibase
  doi.org/10.1016/j.xphs.2018.07.002} {\bibfield  {journal} {\bibinfo
  {journal} {J. Pharm. Sci.}\ }\textbf {\bibinfo {volume} {108}},\ \bibinfo
  {pages} {399--415} (\bibinfo {year} {2019})}\BibitemShut {NoStop}%
\bibitem [{\citenamefont {Schomberg}\ \emph {et~al.}(2021)\citenamefont
  {Schomberg}, \citenamefont {Diener}, \citenamefont {Wünsch}, \citenamefont
  {Finke},\ and\ \citenamefont {Kwade}}]{schomberg:2021}%
  \BibitemOpen
  \bibfield  {author} {\bibinfo {author} {\bibfnamefont {A.~K.}\ \bibnamefont
  {Schomberg}}, \bibinfo {author} {\bibfnamefont {A.}~\bibnamefont {Diener}},
  \bibinfo {author} {\bibfnamefont {I.}~\bibnamefont {Wünsch}}, \bibinfo
  {author} {\bibfnamefont {J.~H.}\ \bibnamefont {Finke}}, \ and\ \bibinfo
  {author} {\bibfnamefont {A.}~\bibnamefont {Kwade}},\ }\bibfield  {title}
  {\enquote {\bibinfo {title} {The use of {X}-ray microtomography to
  investigate the microstructure of pharmaceutical tablets: Potentials and
  comparison to common physical methods},}\ }\href {\doibase
  10.1016/j.ijpx.2021.100090} {\bibfield  {journal} {\bibinfo  {journal}
  {International Journal of Pharmaceutics: X}\ }\textbf {\bibinfo {volume}
  {3}},\ \bibinfo {pages} {100090} (\bibinfo {year} {2021})}\BibitemShut
  {NoStop}%
\bibitem [{\citenamefont {Yost}\ \emph {et~al.}(2019)\citenamefont {Yost},
  \citenamefont {Chalus}, \citenamefont {Zhang}, \citenamefont {Peter},\ and\
  \citenamefont {Narang}}]{yost:2019}%
  \BibitemOpen
  \bibfield  {author} {\bibinfo {author} {\bibfnamefont {E.}~\bibnamefont
  {Yost}}, \bibinfo {author} {\bibfnamefont {P.}~\bibnamefont {Chalus}},
  \bibinfo {author} {\bibfnamefont {S.}~\bibnamefont {Zhang}}, \bibinfo
  {author} {\bibfnamefont {S.}~\bibnamefont {Peter}}, \ and\ \bibinfo {author}
  {\bibfnamefont {A.~S.}\ \bibnamefont {Narang}},\ }\bibfield  {title}
  {\enquote {\bibinfo {title} {Quantitative x-ray microcomputed tomography
  assessment of internal tablet defects},}\ }\href {\doibase
  10.1016/j.xphs.2018.12.024} {\bibfield  {journal} {\bibinfo  {journal}
  {Journal of Pharmaceutical Sciences}\ }\textbf {\bibinfo {volume} {108}},\
  \bibinfo {pages} {1818--1830} (\bibinfo {year} {2019})}\BibitemShut {NoStop}%
\bibitem [{\citenamefont {Sondej}\ \emph {et~al.}(2015)\citenamefont {Sondej},
  \citenamefont {Bück}, \citenamefont {Koslowsky}, \citenamefont {Bachmann},
  \citenamefont {Jacob},\ and\ \citenamefont {Tsotsas}}]{sondej:2015}%
  \BibitemOpen
  \bibfield  {author} {\bibinfo {author} {\bibfnamefont {F.}~\bibnamefont
  {Sondej}}, \bibinfo {author} {\bibfnamefont {A.}~\bibnamefont {Bück}},
  \bibinfo {author} {\bibfnamefont {K.}~\bibnamefont {Koslowsky}}, \bibinfo
  {author} {\bibfnamefont {P.}~\bibnamefont {Bachmann}}, \bibinfo {author}
  {\bibfnamefont {M.}~\bibnamefont {Jacob}}, \ and\ \bibinfo {author}
  {\bibfnamefont {E.}~\bibnamefont {Tsotsas}},\ }\bibfield  {title} {\enquote
  {\bibinfo {title} {Investigation of coating layer morphology by
  micro-computed {X}-ray tomography},}\ }\href {\doibase
  10.1016/j.powtec.2014.12.050} {\bibfield  {journal} {\bibinfo  {journal}
  {Powder Technology}\ }\textbf {\bibinfo {volume} {273}},\ \bibinfo {pages}
  {165--175} (\bibinfo {year} {2015})}\BibitemShut {NoStop}%
\bibitem [{\citenamefont {Zeitler}\ and\ \citenamefont
  {Gladden}(2009)}]{zeitler:2009}%
  \BibitemOpen
  \bibfield  {author} {\bibinfo {author} {\bibfnamefont {J.~A.}\ \bibnamefont
  {Zeitler}}\ and\ \bibinfo {author} {\bibfnamefont {L.~F.}\ \bibnamefont
  {Gladden}},\ }\bibfield  {title} {\enquote {\bibinfo {title} {In-vitro
  tomography and non-destructive imaging at depth of pharmaceutical solid
  dosage forms},}\ }\href {\doibase 10.1016/j.ejpb.2008.08.012} {\bibfield
  {journal} {\bibinfo  {journal} {European Journal of Pharmaceutics and
  Biopharmaceutics}\ }\textbf {\bibinfo {volume} {71}},\ \bibinfo {pages}
  {2--22} (\bibinfo {year} {2009})}\BibitemShut {NoStop}%
\bibitem [{\citenamefont {Hancock}\ and\ \citenamefont
  {Mullarney}(2005)}]{hancock:2005}%
  \BibitemOpen
  \bibfield  {author} {\bibinfo {author} {\bibfnamefont {B.~C.}\ \bibnamefont
  {Hancock}}\ and\ \bibinfo {author} {\bibfnamefont {M.~P.}\ \bibnamefont
  {Mullarney}},\ }\bibfield  {title} {\enquote {\bibinfo {title} {X-ray
  microtomography of solid dosage forms},}\ }\href@noop {} {\bibfield
  {journal} {\bibinfo  {journal} {Pharmaceutical Technology}\ }\textbf
  {\bibinfo {volume} {29}},\ \bibinfo {pages} {92--100} (\bibinfo {year}
  {2005})}\BibitemShut {NoStop}%
\bibitem [{\citenamefont {Gajjar}\ \emph {et~al.}(2020)\citenamefont {Gajjar},
  \citenamefont {Styliari}, \citenamefont {Nguyen}, \citenamefont {Carr},
  \citenamefont {Chen}, \citenamefont {Elliott}, \citenamefont {Hammond},
  \citenamefont {Burnett}, \citenamefont {Roberts}, \citenamefont {Withers},\
  and\ \citenamefont {Murnane}}]{gajjar:2020}%
  \BibitemOpen
  \bibfield  {author} {\bibinfo {author} {\bibfnamefont {P.}~\bibnamefont
  {Gajjar}}, \bibinfo {author} {\bibfnamefont {I.~D.}\ \bibnamefont
  {Styliari}}, \bibinfo {author} {\bibfnamefont {T.~T.~H.}\ \bibnamefont
  {Nguyen}}, \bibinfo {author} {\bibfnamefont {J.}~\bibnamefont {Carr}},
  \bibinfo {author} {\bibfnamefont {X.}~\bibnamefont {Chen}}, \bibinfo {author}
  {\bibfnamefont {J.~A.}\ \bibnamefont {Elliott}}, \bibinfo {author}
  {\bibfnamefont {R.~B.}\ \bibnamefont {Hammond}}, \bibinfo {author}
  {\bibfnamefont {T.~L.}\ \bibnamefont {Burnett}}, \bibinfo {author}
  {\bibfnamefont {K.}~\bibnamefont {Roberts}}, \bibinfo {author} {\bibfnamefont
  {P.~J.}\ \bibnamefont {Withers}}, \ and\ \bibinfo {author} {\bibfnamefont
  {D.}~\bibnamefont {Murnane}},\ }\bibfield  {title} {\enquote {\bibinfo
  {title} {3d characterisation of dry powder inhaler formulations: {D}eveloping
  {X}-ray micro computed tomography approaches},}\ }\href {\doibase
  10.1016/j.ejpb.2020.02.013} {\bibfield  {journal} {\bibinfo  {journal}
  {European Journal of Pharmaceutics and Biopharmaceutics}\ }\textbf {\bibinfo
  {volume} {151}},\ \bibinfo {pages} {32--44} (\bibinfo {year}
  {2020})}\BibitemShut {NoStop}%
\bibitem [{\citenamefont {Wenzel}\ \emph {et~al.}(2021)\citenamefont {Wenzel},
  \citenamefont {Sack}, \citenamefont {Müller}, \citenamefont {P\"oschel},
  \citenamefont {Schuldt-Lieb},\ and\ \citenamefont {Gieseler}}]{wenzel:2021}%
  \BibitemOpen
  \bibfield  {author} {\bibinfo {author} {\bibfnamefont {T.}~\bibnamefont
  {Wenzel}}, \bibinfo {author} {\bibfnamefont {A.}~\bibnamefont {Sack}},
  \bibinfo {author} {\bibfnamefont {P.}~\bibnamefont {Müller}}, \bibinfo
  {author} {\bibfnamefont {T.}~\bibnamefont {P\"oschel}}, \bibinfo {author}
  {\bibfnamefont {S.}~\bibnamefont {Schuldt-Lieb}}, \ and\ \bibinfo {author}
  {\bibfnamefont {H.}~\bibnamefont {Gieseler}},\ }\bibfield  {title} {\enquote
  {\bibinfo {title} {{Stability of freeze-dried products subjected to
  microcomputed tomography radiation doses}},}\ }\href {\doibase
  10.1093/jpp/rgaa004} {\bibfield  {journal} {\bibinfo  {journal} {Journal of
  Pharmacy and Pharmacology}\ }\textbf {\bibinfo {volume} {73}},\ \bibinfo
  {pages} {212--220} (\bibinfo {year} {2021})}\BibitemShut {NoStop}%
\bibitem [{\citenamefont {Stärtzel}\ \emph
  {et~al.}(2015{\natexlab{b}})\citenamefont {Stärtzel}, \citenamefont
  {Gieseler}, \citenamefont {Gieseler}, \citenamefont {Abdul-Fattah},
  \citenamefont {Adler}, \citenamefont {Mahler},\ and\ \citenamefont
  {Goldbach}}]{startzel:2015a}%
  \BibitemOpen
  \bibfield  {author} {\bibinfo {author} {\bibfnamefont {P.}~\bibnamefont
  {Stärtzel}}, \bibinfo {author} {\bibfnamefont {H.}~\bibnamefont {Gieseler}},
  \bibinfo {author} {\bibfnamefont {M.}~\bibnamefont {Gieseler}}, \bibinfo
  {author} {\bibfnamefont {A.M.}\ \bibnamefont {Abdul-Fattah}}, \bibinfo
  {author} {\bibfnamefont {M.}~\bibnamefont {Adler}}, \bibinfo {author}
  {\bibfnamefont {H.C.}\ \bibnamefont {Mahler}}, \ and\ \bibinfo {author}
  {\bibfnamefont {P.}~\bibnamefont {Goldbach}},\ }\bibfield  {title} {\enquote
  {\bibinfo {title} {Freeze drying of l-arginine/sucrose-based protein
  formulations, part 1: Influence of formulation and arginine counter ion on
  the critical formulation temperature, product performance and protein
  stability},}\ }\href {\doibase 10.1002/jps.24501} {\bibfield  {journal}
  {\bibinfo  {journal} {J. Pharm. Sci.}\ }\textbf {\bibinfo {volume} {104}},\
  \bibinfo {pages} {2345--58} (\bibinfo {year}
  {2015}{\natexlab{b}})}\BibitemShut {NoStop}%
\bibitem [{\citenamefont {Szegedy}\ \emph {et~al.}(2015)\citenamefont
  {Szegedy}, \citenamefont {Liu}, \citenamefont {Y.}, \citenamefont {Sermanet},
  \citenamefont {Reed}, \citenamefont {Anguelov}, \citenamefont {Erhan},
  \citenamefont {Vanhoucke},\ and\ \citenamefont {Rabinovich}}]{inception}%
  \BibitemOpen
  \bibfield  {author} {\bibinfo {author} {\bibfnamefont {C.}~\bibnamefont
  {Szegedy}}, \bibinfo {author} {\bibfnamefont {W.}~\bibnamefont {Liu}},
  \bibinfo {author} {\bibfnamefont {Jia.}\ \bibnamefont {Y.}}, \bibinfo
  {author} {\bibfnamefont {P.}~\bibnamefont {Sermanet}}, \bibinfo {author}
  {\bibfnamefont {S.}~\bibnamefont {Reed}}, \bibinfo {author} {\bibfnamefont
  {D.}~\bibnamefont {Anguelov}}, \bibinfo {author} {\bibfnamefont
  {D.}~\bibnamefont {Erhan}}, \bibinfo {author} {\bibfnamefont
  {V.}~\bibnamefont {Vanhoucke}}, \ and\ \bibinfo {author} {\bibfnamefont
  {A.}~\bibnamefont {Rabinovich}},\ }\bibfield  {title} {\enquote {\bibinfo
  {title} {Going deeper with convolutions},}\ }in\ \href {\doibase
  10.1109/CVPR.2015.7298594} {\emph {\bibinfo {booktitle} {2015 IEEE Conference
  on Computer Vision and Pattern Recognition (CVPR)}}}\ (\bibinfo {year}
  {2015})\ pp.\ \bibinfo {pages} {1--9}\BibitemShut {NoStop}%
\bibitem [{\citenamefont {Sam}\ \emph {et~al.}(2019)\citenamefont {Sam},
  \citenamefont {Kamardin}, \citenamefont {Sjarif},\ and\ \citenamefont
  {Mohamed}}]{Sam2019OfflineSVCGPT}%
  \BibitemOpen
  \bibfield  {author} {\bibinfo {author} {\bibfnamefont {S.~M.}\ \bibnamefont
  {Sam}}, \bibinfo {author} {\bibfnamefont {K.}~\bibnamefont {Kamardin}},
  \bibinfo {author} {\bibfnamefont {S.~M.}\ \bibnamefont {Sjarif}}, \ and\
  \bibinfo {author} {\bibfnamefont {N.}~\bibnamefont {Mohamed}},\ }\bibfield
  {title} {\enquote {\bibinfo {title} {Offline signature verification using
  deep learning convolutional neural network ({CNN}) architectures
  {G}oog{L}e{N}et {I}nception-v1 and {I}nception-v3},}\ }\href {\doibase
  10.1016/j.procs.2019.12.108} {\bibfield  {journal} {\bibinfo  {journal}
  {Procedia Computer Science}\ }\textbf {\bibinfo {volume} {163}},\ \bibinfo
  {pages} {84--91} (\bibinfo {year} {2019})}\BibitemShut {NoStop}%
\bibitem [{\citenamefont {Xia}\ \emph {et~al.}(2017)\citenamefont {Xia},
  \citenamefont {Xu},\ and\ \citenamefont {Nan}}]{Xia2017Inceptionv3FFCGPT}%
  \BibitemOpen
  \bibfield  {author} {\bibinfo {author} {\bibfnamefont {X.}~\bibnamefont
  {Xia}}, \bibinfo {author} {\bibfnamefont {C.}~\bibnamefont {Xu}}, \ and\
  \bibinfo {author} {\bibfnamefont {B.}~\bibnamefont {Nan}},\ }\bibfield
  {title} {\enquote {\bibinfo {title} {Inception-v3 for flower
  classification},}\ }in\ \href {\doibase 10.1109/ICIVC.2017.7984661} {\emph
  {\bibinfo {booktitle} {2017 2nd International Conference on Image, Vision and
  Computing (ICIVC)}}}\ (\bibinfo {organization} {IEEE},\ \bibinfo {year}
  {2017})\BibitemShut {NoStop}%
\bibitem [{\citenamefont {Wang}\ \emph {et~al.}(2019)\citenamefont {Wang},
  \citenamefont {Chen}, \citenamefont {Hao}, \citenamefont {Liu}, \citenamefont
  {Zeng},\ and\ \citenamefont {Chen}}]{Wang2019PulmonaryICCGPT}%
  \BibitemOpen
  \bibfield  {author} {\bibinfo {author} {\bibfnamefont {C}~\bibnamefont
  {Wang}}, \bibinfo {author} {\bibfnamefont {D.}~\bibnamefont {Chen}}, \bibinfo
  {author} {\bibfnamefont {L.}~\bibnamefont {Hao}}, \bibinfo {author}
  {\bibfnamefont {X.}~\bibnamefont {Liu}}, \bibinfo {author} {\bibfnamefont
  {Y.}~\bibnamefont {Zeng}}, \ and\ \bibinfo {author} {\bibfnamefont
  {J.}~\bibnamefont {Chen}},\ }\bibfield  {title} {\enquote {\bibinfo {title}
  {Pulmonary image classification based on {I}nception-v3 transfer learning
  model},}\ }in\ \href {\doibase 10.1109/SIPROCESS.2019.8861312} {\emph
  {\bibinfo {booktitle} {2019 IEEE 4th International Conference on Signal and
  Image Processing (ICSIP)}}}\ (\bibinfo {organization} {IEEE},\ \bibinfo
  {year} {2019})\BibitemShut {NoStop}%
\bibitem [{\citenamefont {Yang}\ \emph {et~al.}(2020)\citenamefont {Yang},
  \citenamefont {Zhang}, \citenamefont {Dai},\ and\ \citenamefont
  {Pan}}]{transfer}%
  \BibitemOpen
  \bibfield  {author} {\bibinfo {author} {\bibfnamefont {Qiang}\ \bibnamefont
  {Yang}}, \bibinfo {author} {\bibfnamefont {Yu}~\bibnamefont {Zhang}},
  \bibinfo {author} {\bibfnamefont {Wenyuan}\ \bibnamefont {Dai}}, \ and\
  \bibinfo {author} {\bibfnamefont {Sinno~Jialin}\ \bibnamefont {Pan}},\
  }\href@noop {} {\emph {\bibinfo {title} {Transfer Learning}}}\ (\bibinfo
  {publisher} {Cambridge University Press},\ \bibinfo {year}
  {2020})\BibitemShut {NoStop}%
\bibitem [{\citenamefont {Lu}\ \emph {et~al.}(2015)\citenamefont {Lu},
  \citenamefont {Behbood}, \citenamefont {Hao}, \citenamefont {Zuo},
  \citenamefont {Xue},\ and\ \citenamefont {Zhang}}]{Lu2015TransferLUCGPT}%
  \BibitemOpen
  \bibfield  {author} {\bibinfo {author} {\bibfnamefont {J.}~\bibnamefont
  {Lu}}, \bibinfo {author} {\bibfnamefont {V.}~\bibnamefont {Behbood}},
  \bibinfo {author} {\bibfnamefont {P.}~\bibnamefont {Hao}}, \bibinfo {author}
  {\bibfnamefont {H.}~\bibnamefont {Zuo}}, \bibinfo {author} {\bibfnamefont
  {S.}~\bibnamefont {Xue}}, \ and\ \bibinfo {author} {\bibfnamefont
  {G.}~\bibnamefont {Zhang}},\ }\bibfield  {title} {\enquote {\bibinfo {title}
  {Transfer learning using computational intelligence: {A} survey},}\ }\href
  {\doibase 10.1016/j.knosys.2015.01.010} {\bibfield  {journal} {\bibinfo
  {journal} {Knowledge-Based Systems}\ }\textbf {\bibinfo {volume} {80}},\
  \bibinfo {pages} {14--23} (\bibinfo {year} {2015})}\BibitemShut {NoStop}%
\bibitem [{\citenamefont {Zhuang}\ \emph {et~al.}(2020)\citenamefont {Zhuang},
  \citenamefont {Qi}, \citenamefont {Duan}, \citenamefont {Xi}, \citenamefont
  {Zhu}, \citenamefont {Zhu}, \citenamefont {Xiong},\ and\ \citenamefont
  {He}}]{Zhuang2020ComprehensiveSOCGPT}%
  \BibitemOpen
  \bibfield  {author} {\bibinfo {author} {\bibfnamefont {F.}~\bibnamefont
  {Zhuang}}, \bibinfo {author} {\bibfnamefont {Z.}~\bibnamefont {Qi}}, \bibinfo
  {author} {\bibfnamefont {K.}~\bibnamefont {Duan}}, \bibinfo {author}
  {\bibfnamefont {D.}~\bibnamefont {Xi}}, \bibinfo {author} {\bibfnamefont
  {Y.}~\bibnamefont {Zhu}}, \bibinfo {author} {\bibfnamefont {H.}~\bibnamefont
  {Zhu}}, \bibinfo {author} {\bibfnamefont {H.}~\bibnamefont {Xiong}}, \ and\
  \bibinfo {author} {\bibfnamefont {Q.}~\bibnamefont {He}},\ }\bibfield
  {title} {\enquote {\bibinfo {title} {A comprehensive survey on transfer
  learning},}\ }\href {\doibase 10.1109/JPROC.2020.3004555} {\bibfield
  {journal} {\bibinfo  {journal} {Proceedings of the IEEE}\ }\textbf {\bibinfo
  {volume} {109}},\ \bibinfo {pages} {43--76} (\bibinfo {year}
  {2020})}\BibitemShut {NoStop}%
\bibitem [{\citenamefont {Gao}\ and\ \citenamefont
  {Mosalam}(2018)}]{Gao2018DeepTLCGPT}%
  \BibitemOpen
  \bibfield  {author} {\bibinfo {author} {\bibfnamefont {Y.}~\bibnamefont
  {Gao}}\ and\ \bibinfo {author} {\bibfnamefont {K.~M.}\ \bibnamefont
  {Mosalam}},\ }\bibfield  {title} {\enquote {\bibinfo {title} {Deep transfer
  learning for image‐based structural damage recognition},}\ }\href {\doibase
  10.1111/mice.12363} {\bibfield  {journal} {\bibinfo  {journal}
  {Computer‐Aided Civil and Infrastructure Engineering}\ }\textbf {\bibinfo
  {volume} {33}},\ \bibinfo {pages} {748--768} (\bibinfo {year}
  {2018})}\BibitemShut {NoStop}%
\bibitem [{\citenamefont {Ghafoorian}\ \emph {et~al.}(2017)\citenamefont
  {Ghafoorian}, \citenamefont {Mehrtash}, \citenamefont {Kapur}, \citenamefont
  {Karssemeijer}, \citenamefont {Marchiori}, \citenamefont {Pesteie},
  \citenamefont {Guttmann}, \citenamefont {de~Leeuw}, \citenamefont {Tempany},
  \citenamefont {van Ginneken}, \citenamefont {Fedorov}, \citenamefont
  {Abolmaesumi}, \citenamefont {Platel},\ and\ \citenamefont
  {Wells}}]{Ghafoorian2017TransferLFCGPT}%
  \BibitemOpen
  \bibfield  {author} {\bibinfo {author} {\bibfnamefont {M.}~\bibnamefont
  {Ghafoorian}}, \bibinfo {author} {\bibfnamefont {A.}~\bibnamefont
  {Mehrtash}}, \bibinfo {author} {\bibfnamefont {T.}~\bibnamefont {Kapur}},
  \bibinfo {author} {\bibfnamefont {N.}~\bibnamefont {Karssemeijer}}, \bibinfo
  {author} {\bibfnamefont {E.}~\bibnamefont {Marchiori}}, \bibinfo {author}
  {\bibfnamefont {M.n}\ \bibnamefont {Pesteie}}, \bibinfo {author}
  {\bibfnamefont {C.~R.~G.}\ \bibnamefont {Guttmann}}, \bibinfo {author}
  {\bibfnamefont {F.-E.}\ \bibnamefont {de~Leeuw}}, \bibinfo {author}
  {\bibfnamefont {C.~M.}\ \bibnamefont {Tempany}}, \bibinfo {author}
  {\bibfnamefont {B.}~\bibnamefont {van Ginneken}}, \bibinfo {author}
  {\bibfnamefont {A.}~\bibnamefont {Fedorov}}, \bibinfo {author} {\bibfnamefont
  {P.}~\bibnamefont {Abolmaesumi}}, \bibinfo {author} {\bibfnamefont
  {B.}~\bibnamefont {Platel}}, \ and\ \bibinfo {author} {\bibfnamefont {W.~M.}\
  \bibnamefont {Wells}},\ }\bibfield  {title} {\enquote {\bibinfo {title}
  {Transfer learning for domain adaptation in {MRI}: {A}pplication in brain
  lesion segmentation},}\ }in\ \href {\doibase 10.1007/978-3-319-66179-7_59}
  {\emph {\bibinfo {booktitle} {Medical Image Computing and Computer Assisted
  Intervention -- MICCAI 2017}}},\ \bibinfo {editor} {edited by\ \bibinfo
  {editor} {\bibfnamefont {M.}~\bibnamefont {Descoteaux}}, \bibinfo {editor}
  {\bibfnamefont {L.}~\bibnamefont {Maier-Hein}}, \bibinfo {editor}
  {\bibfnamefont {A.}~\bibnamefont {Franz}}, \bibinfo {editor} {\bibfnamefont
  {P.}~\bibnamefont {Jannin}}, \bibinfo {editor} {\bibfnamefont {D.~L.}\
  \bibnamefont {Collins}}, \ and\ \bibinfo {editor} {\bibfnamefont
  {S.}~\bibnamefont {Duchesne}}}\ (\bibinfo  {publisher} {Springer
  International Publishing},\ \bibinfo {address} {Cham},\ \bibinfo {year}
  {2017})\ pp.\ \bibinfo {pages} {516--524}\BibitemShut {NoStop}%
\bibitem [{\citenamefont {Minaee}\ \emph {et~al.}(2020)\citenamefont {Minaee},
  \citenamefont {Kafieh}, \citenamefont {Sonka}, \citenamefont {Yazdani},\ and\
  \citenamefont {Jamalipour}}]{Minaee2020DeepCOVIDPCCGPT}%
  \BibitemOpen
  \bibfield  {author} {\bibinfo {author} {\bibfnamefont {S.}~\bibnamefont
  {Minaee}}, \bibinfo {author} {\bibfnamefont {R.}~\bibnamefont {Kafieh}},
  \bibinfo {author} {\bibfnamefont {M.}~\bibnamefont {Sonka}}, \bibinfo
  {author} {\bibfnamefont {S.}~\bibnamefont {Yazdani}}, \ and\ \bibinfo
  {author} {\bibfnamefont {G.~H.~R.}\ \bibnamefont {Jamalipour}},\ }\bibfield
  {title} {\enquote {\bibinfo {title} {Deep-{COVID}: {P}redicting {COVID}-19
  from chest {X}-ray images using deep transfer learning},}\ }\href {\doibase
  10.1016/j.media.2020.101794} {\bibfield  {journal} {\bibinfo  {journal}
  {Medical Image Analysis}\ }\textbf {\bibinfo {volume} {65}},\ \bibinfo
  {pages} {101794} (\bibinfo {year} {2020})}\BibitemShut {NoStop}%
\bibitem [{\citenamefont {Nair}\ \emph {et~al.}(2021)\citenamefont {Nair},
  \citenamefont {Mühlbauer}, \citenamefont {Roy},\ and\ \citenamefont
  {Pöschel}}]{faucris.258191566CGPT}%
  \BibitemOpen
  \bibfield  {author} {\bibinfo {author} {\bibfnamefont {P.}~\bibnamefont
  {Nair}}, \bibinfo {author} {\bibfnamefont {S.}~\bibnamefont {Mühlbauer}},
  \bibinfo {author} {\bibfnamefont {S.}~\bibnamefont {Roy}}, \ and\ \bibinfo
  {author} {\bibfnamefont {T.}~\bibnamefont {Pöschel}},\ }\bibfield  {title}
  {\enquote {\bibinfo {title} {{Can} {Minkowski} tensors of a simply connected
  porous microstructure characterize its permeability?}}\ }\href {\doibase
  10.1063/5.0045701} {\bibfield  {journal} {\bibinfo  {journal} {Physics of
  Fluids}\ }\textbf {\bibinfo {volume} {33}} (\bibinfo {year} {2021}),\
  10.1063/5.0045701}\BibitemShut {NoStop}%
\bibitem [{\citenamefont {Deng}\ \emph {et~al.}(2009)\citenamefont {Deng},
  \citenamefont {Dong}, \citenamefont {Socher}, \citenamefont {Li},
  \citenamefont {Li},\ and\ \citenamefont {Fei-Fei}}]{imageNet}%
  \BibitemOpen
  \bibfield  {author} {\bibinfo {author} {\bibfnamefont {J.}~\bibnamefont
  {Deng}}, \bibinfo {author} {\bibfnamefont {W.}~\bibnamefont {Dong}}, \bibinfo
  {author} {\bibfnamefont {R.}~\bibnamefont {Socher}}, \bibinfo {author}
  {\bibfnamefont {L.-J.}\ \bibnamefont {Li}}, \bibinfo {author} {\bibfnamefont
  {K.}~\bibnamefont {Li}}, \ and\ \bibinfo {author} {\bibfnamefont
  {L.}~\bibnamefont {Fei-Fei}},\ }\bibfield  {title} {\enquote {\bibinfo
  {title} {Imagenet: {A} large-scale hierarchical image database},}\ }in\ \href
  {\doibase 10.1109/CVPR.2009.5206848} {\emph {\bibinfo {booktitle} {2009 IEEE
  Conference on Computer Vision and Pattern Recognition}}}\ (\bibinfo {year}
  {2009})\ pp.\ \bibinfo {pages} {248--255}\BibitemShut {NoStop}%
\bibitem [{\citenamefont {Russakovsky}\ \emph {et~al.}(2015)\citenamefont
  {Russakovsky}, \citenamefont {Deng}, \citenamefont {Su}, \citenamefont
  {Krause}, \citenamefont {Satheesh}, \citenamefont {Ma}, \citenamefont
  {Huang}, \citenamefont {Karpathy}, \citenamefont {Khosla}, \citenamefont
  {Bernstein}, \citenamefont {Berg},\ and\ \citenamefont
  {Fei-Fei}}]{Russakovsky2015ImageNetLSCGPT}%
  \BibitemOpen
  \bibfield  {author} {\bibinfo {author} {\bibfnamefont {O.}~\bibnamefont
  {Russakovsky}}, \bibinfo {author} {\bibfnamefont {J.}~\bibnamefont {Deng}},
  \bibinfo {author} {\bibfnamefont {H.}~\bibnamefont {Su}}, \bibinfo {author}
  {\bibfnamefont {J.}~\bibnamefont {Krause}}, \bibinfo {author} {\bibfnamefont
  {S.}~\bibnamefont {Satheesh}}, \bibinfo {author} {\bibfnamefont
  {S.}~\bibnamefont {Ma}}, \bibinfo {author} {\bibfnamefont {Z.}~\bibnamefont
  {Huang}}, \bibinfo {author} {\bibfnamefont {A.}~\bibnamefont {Karpathy}},
  \bibinfo {author} {\bibfnamefont {A.}~\bibnamefont {Khosla}}, \bibinfo
  {author} {\bibfnamefont {M.}~\bibnamefont {Bernstein}}, \bibinfo {author}
  {\bibfnamefont {A.~C.}\ \bibnamefont {Berg}}, \ and\ \bibinfo {author}
  {\bibfnamefont {L.}~\bibnamefont {Fei-Fei}},\ }\bibfield  {title} {\enquote
  {\bibinfo {title} {Image{N}et large scale visual recognition challenge},}\
  }\href {\doibase 10.1007/s11263-015-0816-y} {\bibfield  {journal} {\bibinfo
  {journal} {International Journal of Computer Vision}\ }\textbf {\bibinfo
  {volume} {115}},\ \bibinfo {pages} {211--252} (\bibinfo {year}
  {2015})}\BibitemShut {NoStop}%
\bibitem [{\citenamefont {Kornblith}\ \emph {et~al.}(2019)\citenamefont
  {Kornblith}, \citenamefont {Shlens},\ and\ \citenamefont
  {Le}}]{Kornblith2019DoBICGPT}%
  \BibitemOpen
  \bibfield  {author} {\bibinfo {author} {\bibfnamefont {S.}~\bibnamefont
  {Kornblith}}, \bibinfo {author} {\bibfnamefont {J.}~\bibnamefont {Shlens}}, \
  and\ \bibinfo {author} {\bibfnamefont {Q.~V.}\ \bibnamefont {Le}},\
  }\bibfield  {title} {\enquote {\bibinfo {title} {Do better image{N}et models
  transfer better?}}\ }in\ \href
  {http://openaccess.thecvf.com/content_CVPR_2019/html/Kornblith_Do_Better_ImageNet_Models_Transfer_Better_CVPR_2019_paper.html}
  {\emph {\bibinfo {booktitle} {Proceedings of the IEEE/CVF Conference on
  Computer Vision and Pattern Recognition (CVPR)}}}\ (\bibinfo {year}
  {2019})\BibitemShut {NoStop}%
\bibitem [{\citenamefont {Krizhevsky}\ \emph {et~al.}(2017)\citenamefont
  {Krizhevsky}, \citenamefont {Sutskever},\ and\ \citenamefont
  {Hinton}}]{Krizhevsky2017ImageNetCWCGPT}%
  \BibitemOpen
  \bibfield  {author} {\bibinfo {author} {\bibfnamefont {A.}~\bibnamefont
  {Krizhevsky}}, \bibinfo {author} {\bibfnamefont {I.}~\bibnamefont
  {Sutskever}}, \ and\ \bibinfo {author} {\bibfnamefont {G.~E.}\ \bibnamefont
  {Hinton}},\ }\bibfield  {title} {\enquote {\bibinfo {title} {Image{N}et
  classification with deep convolutional neural networks},}\ }\href {\doibase
  10.1145/3065386} {\bibfield  {journal} {\bibinfo  {journal} {Communications
  of the ACM}\ }\textbf {\bibinfo {volume} {60}},\ \bibinfo {pages} {84--90}
  (\bibinfo {year} {2017})}\BibitemShut {NoStop}%
\bibitem [{\citenamefont {Murphy}(2012)}]{murphy:2012}%
  \BibitemOpen
  \bibfield  {author} {\bibinfo {author} {\bibfnamefont {K.}~\bibnamefont
  {Murphy}},\ }\enquote {\bibinfo {title} {Machine learning: a probabilistic
  perspective (adaptive computation and machine learning series)},}\ \
  (\bibinfo  {publisher} {MIT Press},\ \bibinfo {year} {2012})\BibitemShut
  {NoStop}%
\bibitem [{\citenamefont {Li}\ and\ \citenamefont
  {Lu}(2021)}]{Li2021MixedCECGPT}%
  \BibitemOpen
  \bibfield  {author} {\bibinfo {author} {\bibfnamefont {H.}~\bibnamefont
  {Li}}\ and\ \bibinfo {author} {\bibfnamefont {W.}~\bibnamefont {Lu}},\
  }\bibfield  {title} {\enquote {\bibinfo {title} {Mixed cross entropy loss for
  neural machine translation},}\ }in\ \href
  {https://proceedings.mlr.press/v139/li21n.html} {\emph {\bibinfo {booktitle}
  {International Conference on Machine Learning}}},\ Vol.\ \bibinfo {volume}
  {139}\ (\bibinfo {year} {2021})\ pp.\ \bibinfo {pages}
  {6242--6251}\BibitemShut {NoStop}%
\bibitem [{\citenamefont {Ho}\ and\ \citenamefont
  {Wookey}(2019)}]{Ho2019TheRWCGPT}%
  \BibitemOpen
  \bibfield  {author} {\bibinfo {author} {\bibfnamefont {Y.}~\bibnamefont
  {Ho}}\ and\ \bibinfo {author} {\bibfnamefont {S.}~\bibnamefont {Wookey}},\
  }\bibfield  {title} {\enquote {\bibinfo {title} {The real-world-weight
  cross-entropy loss function: {M}odeling the costs of mislabeling},}\ }\href
  {\doibase 10.1109/ACCESS.2019.2956748} {\bibfield  {journal} {\bibinfo
  {journal} {IEEE Access}\ }\textbf {\bibinfo {volume} {7}},\ \bibinfo {pages}
  {176073--176082} (\bibinfo {year} {2019})}\BibitemShut {NoStop}%
\bibitem [{Cha()}]{ChatGPT}%
  \BibitemOpen
  \href@noop {} {}\bibinfo {note} {ChatGPT (Large language model).
  \url{https://chat.openai.com}}\BibitemShut {NoStop}%
\end{thebibliography}%

\hspace*{0cm}\fbox{\includegraphics[height=24cm,bb=50 60 570 800,clip] {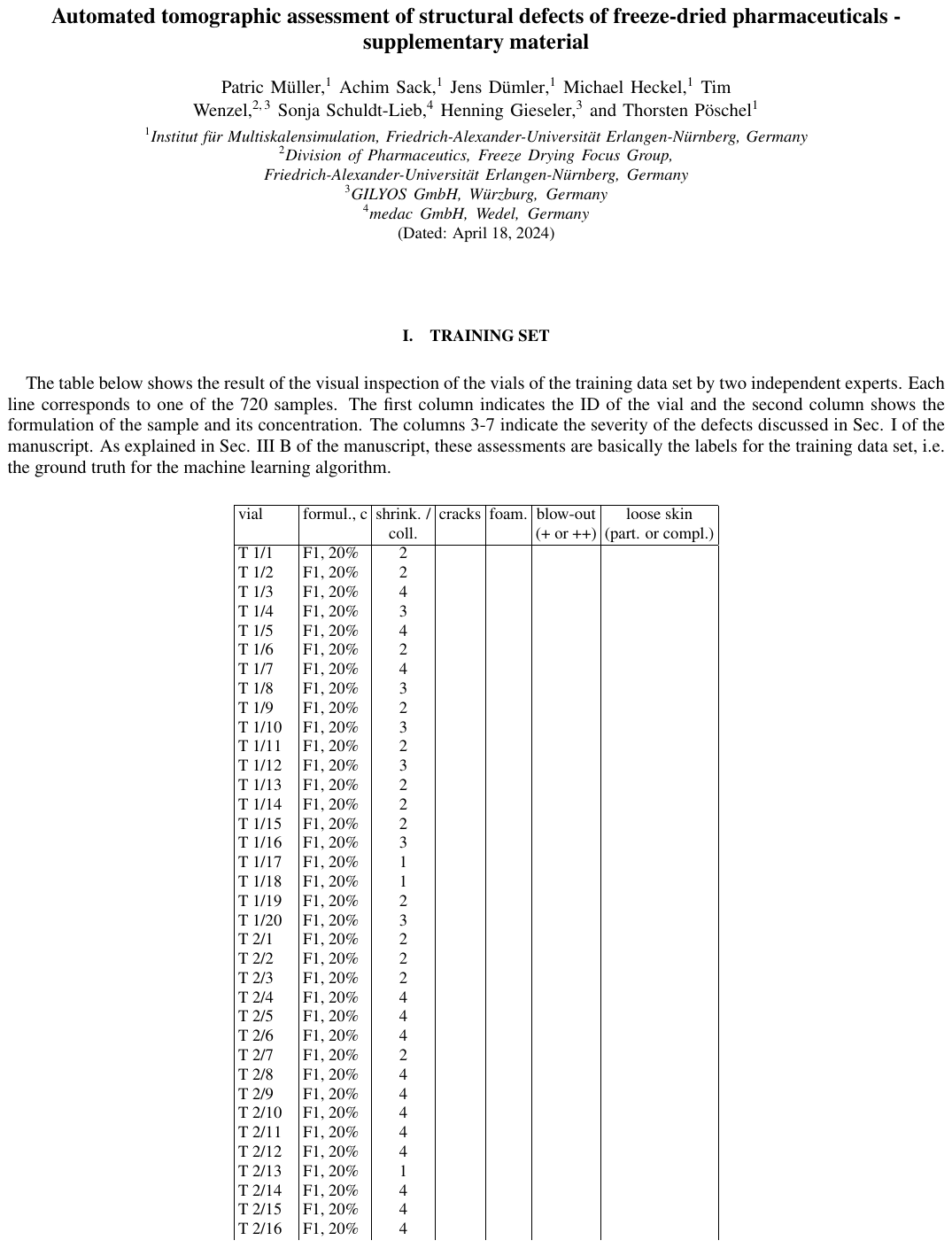}}
\clearpage

\hspace*{0cm}\fbox{\includegraphics[height=24cm,bb=50 60 570 800,clip] {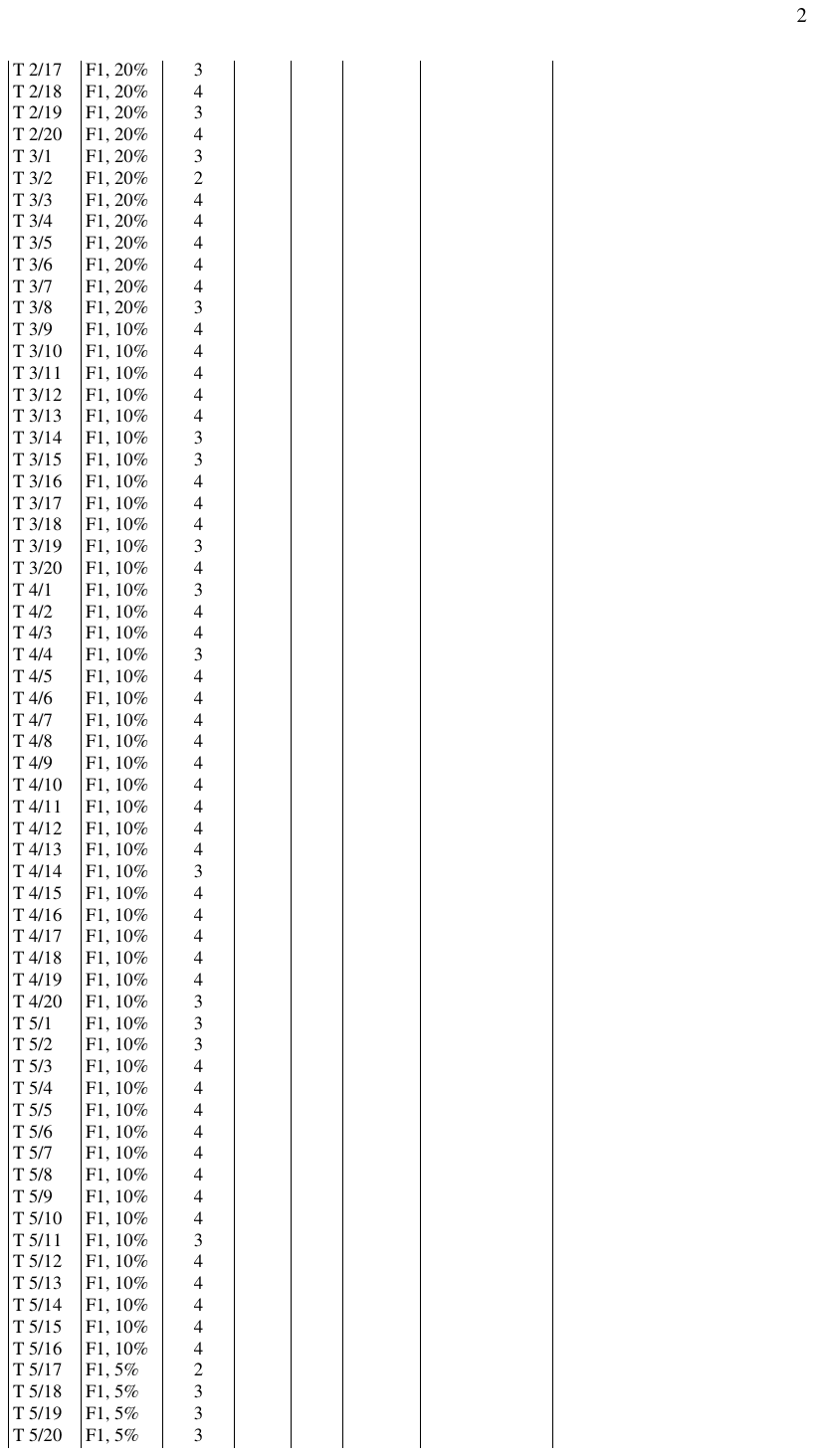}}
\clearpage

\hspace*{0cm}\fbox{\includegraphics[height=24cm,bb=50 60 570 800,clip] {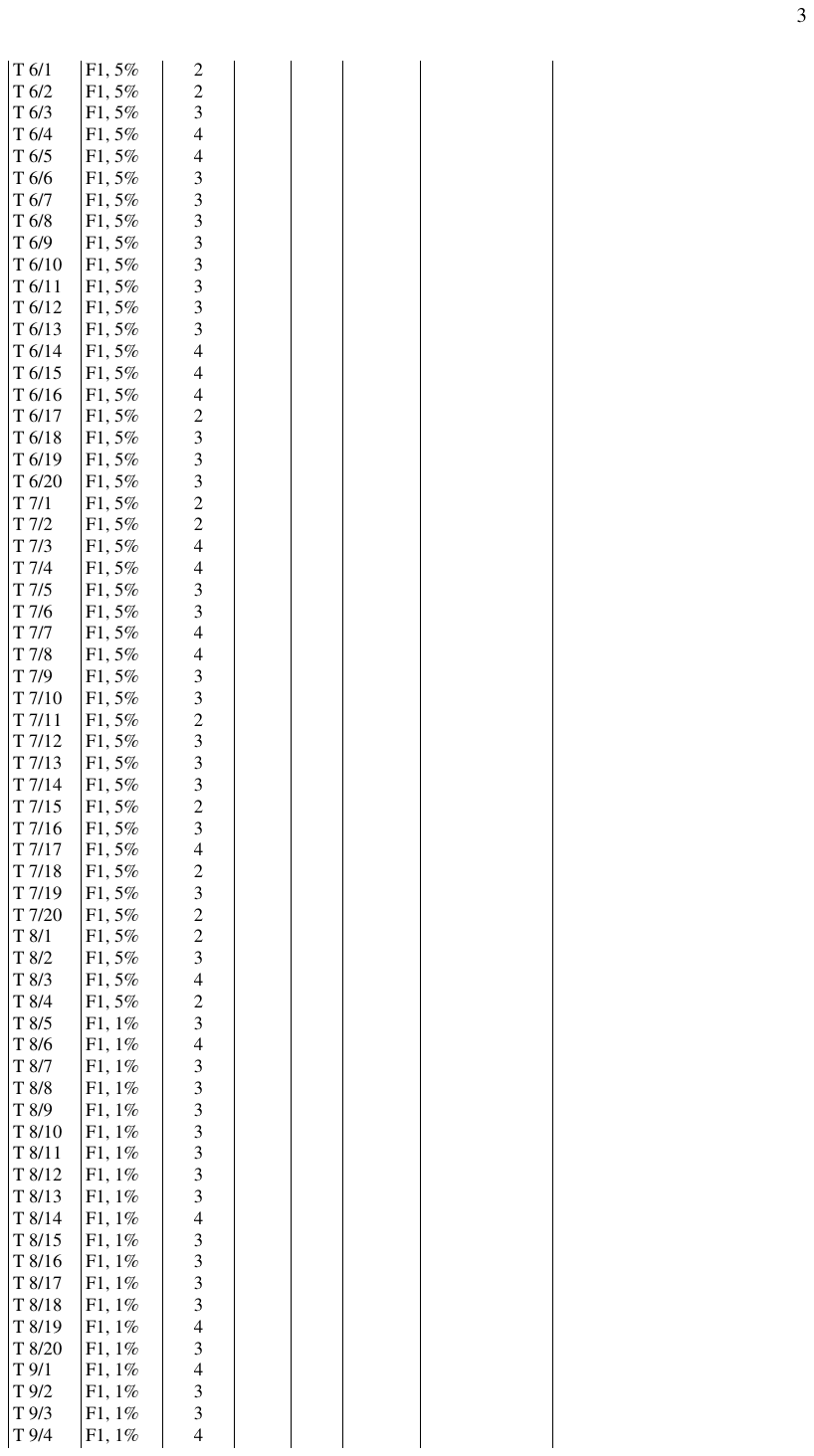}}
\clearpage

\hspace*{0cm}\fbox{\includegraphics[height=24cm,bb=50 60 570 800,clip] {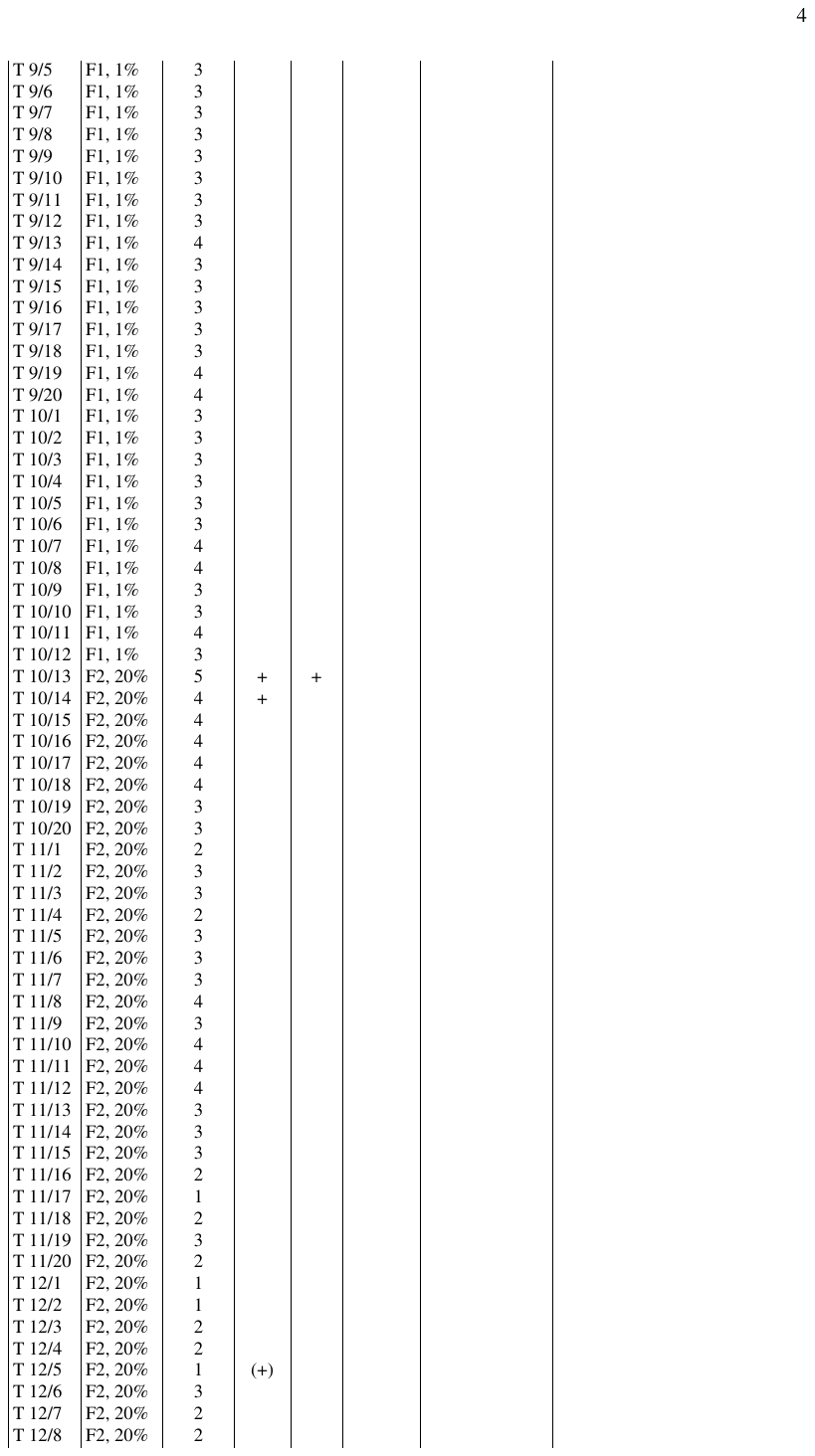}}
\clearpage

\hspace*{0cm}\fbox{\includegraphics[height=24cm,bb=50 60 570 800,clip] {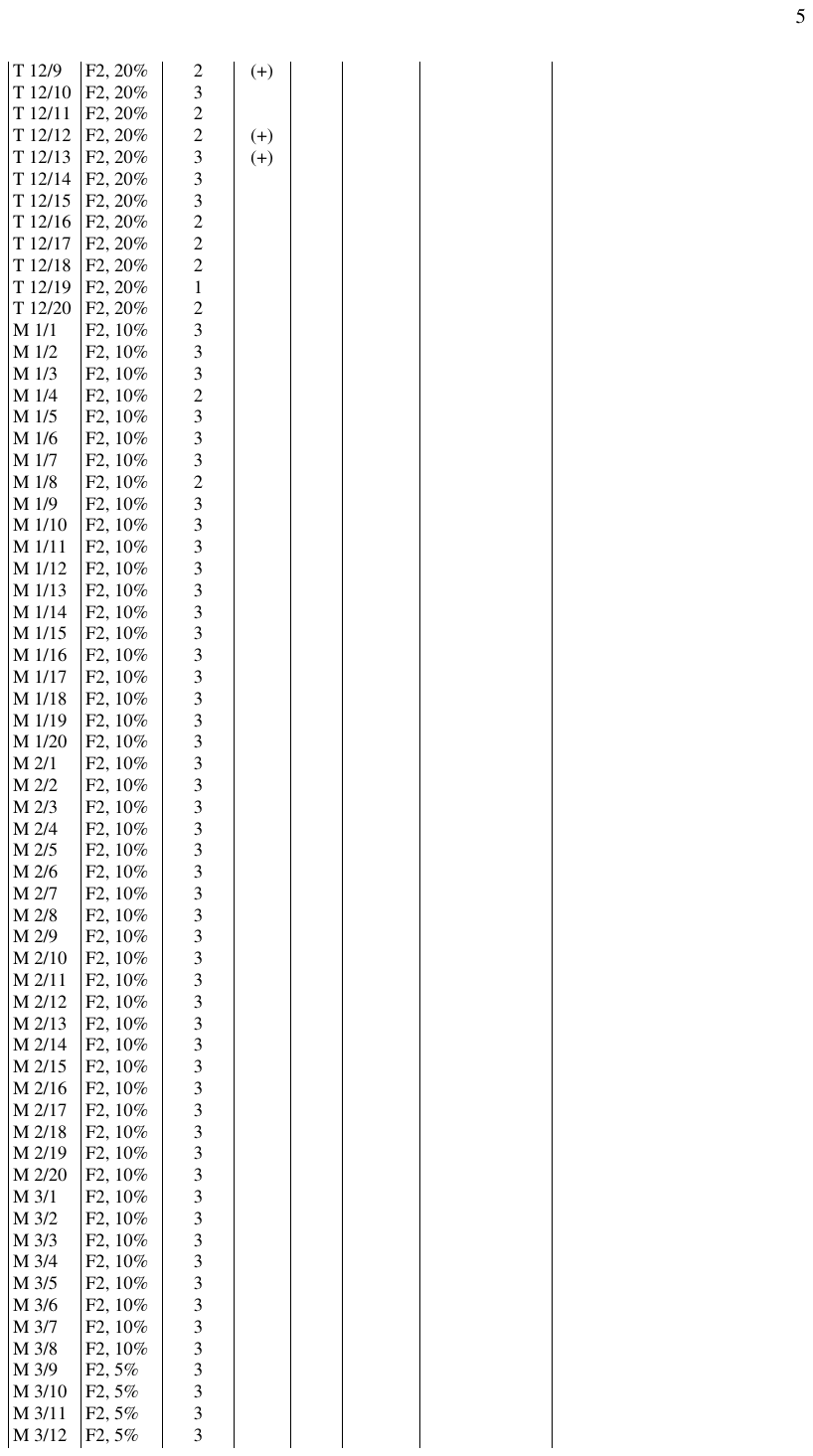}}
\clearpage

\hspace*{0cm}\fbox{\includegraphics[height=24cm,bb=50 60 570 800,clip] {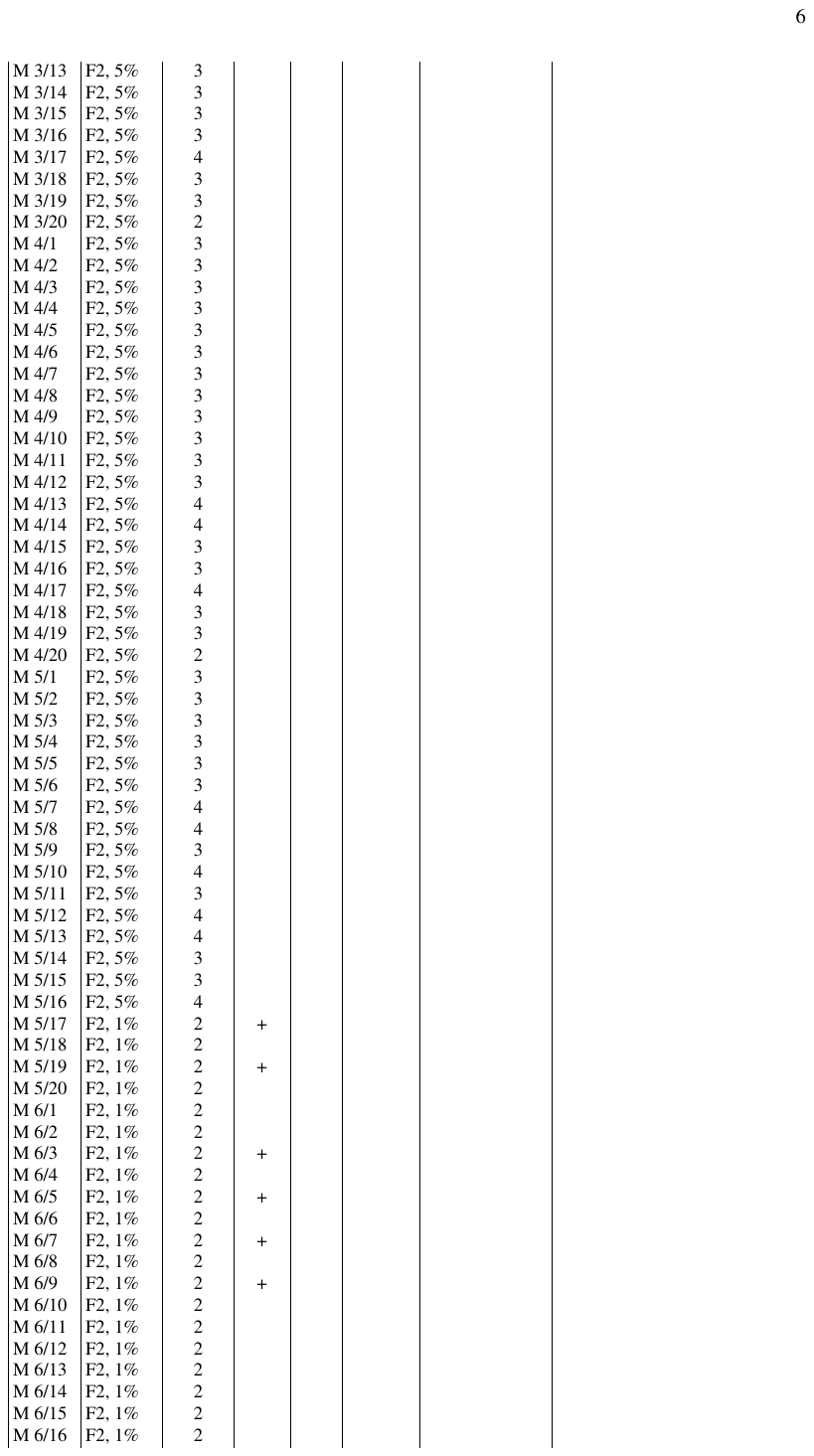}}
\clearpage

\hspace*{0cm}\fbox{\includegraphics[height=24cm,bb=50 60 570 800,clip] {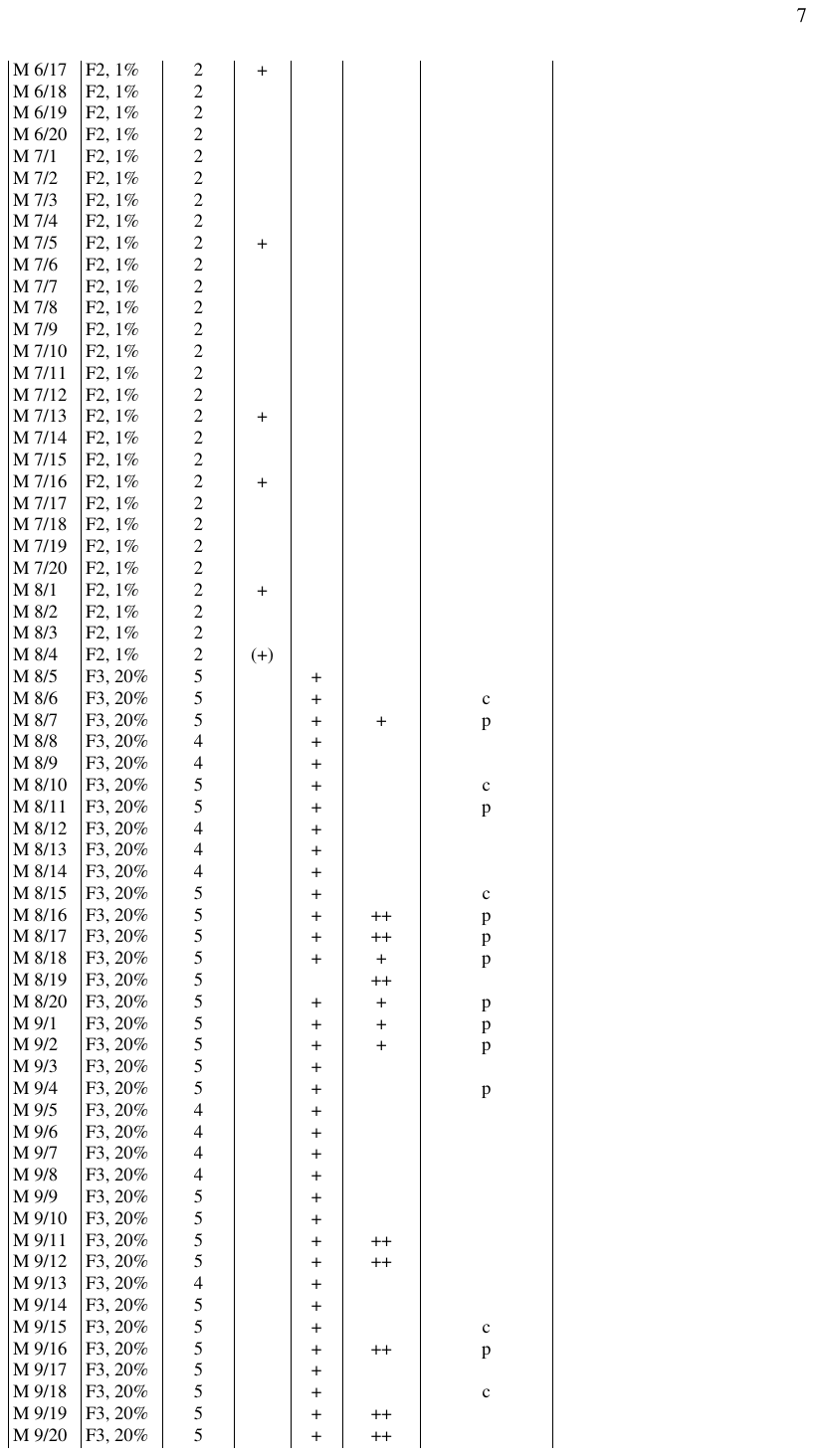}}
\clearpage

\hspace*{0cm}\fbox{\includegraphics[height=24cm,bb=50 60 570 800,clip] {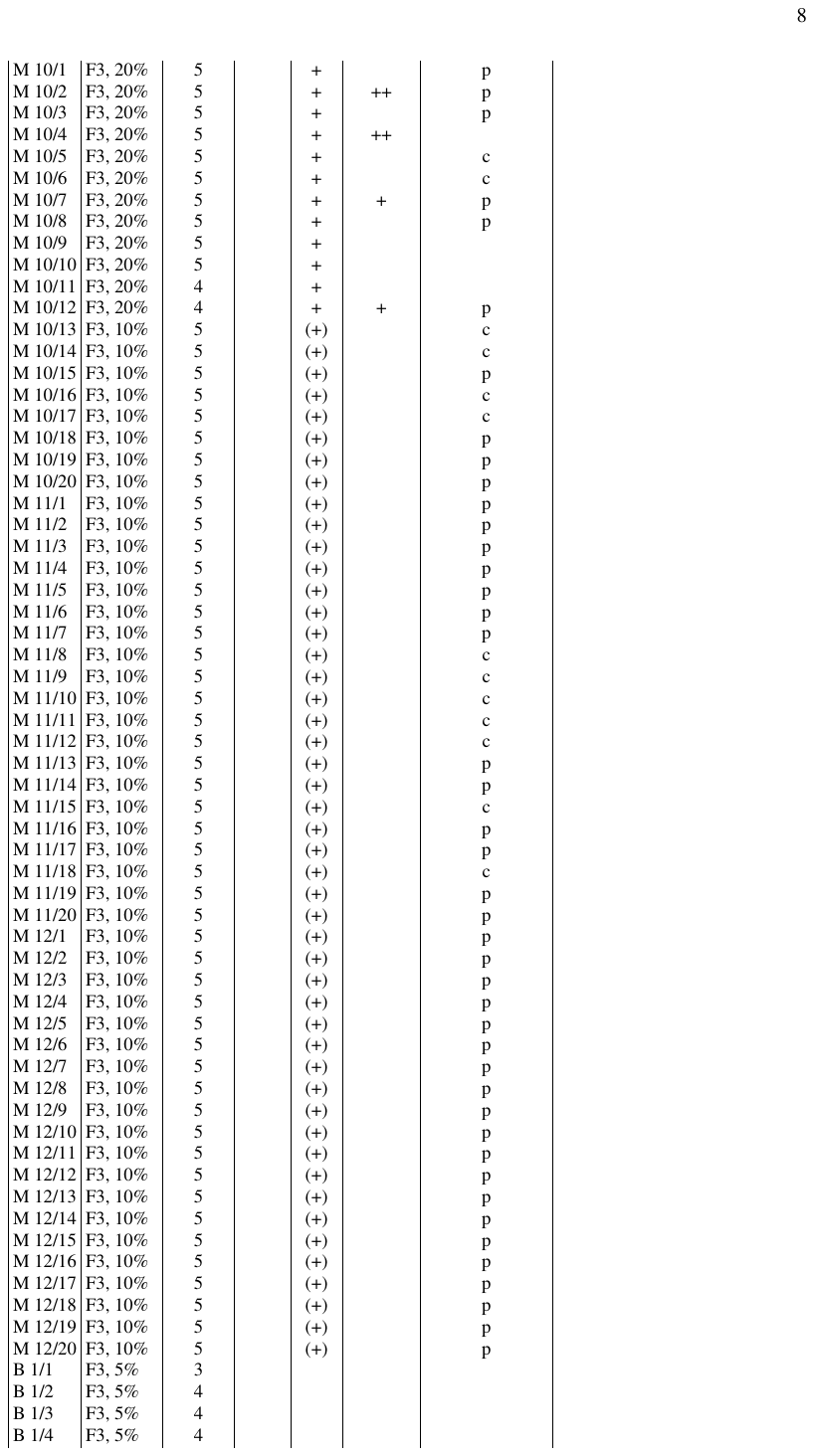}}
\clearpage

\hspace*{0cm}\fbox{\includegraphics[height=24cm,bb=50 60 570 800,clip] {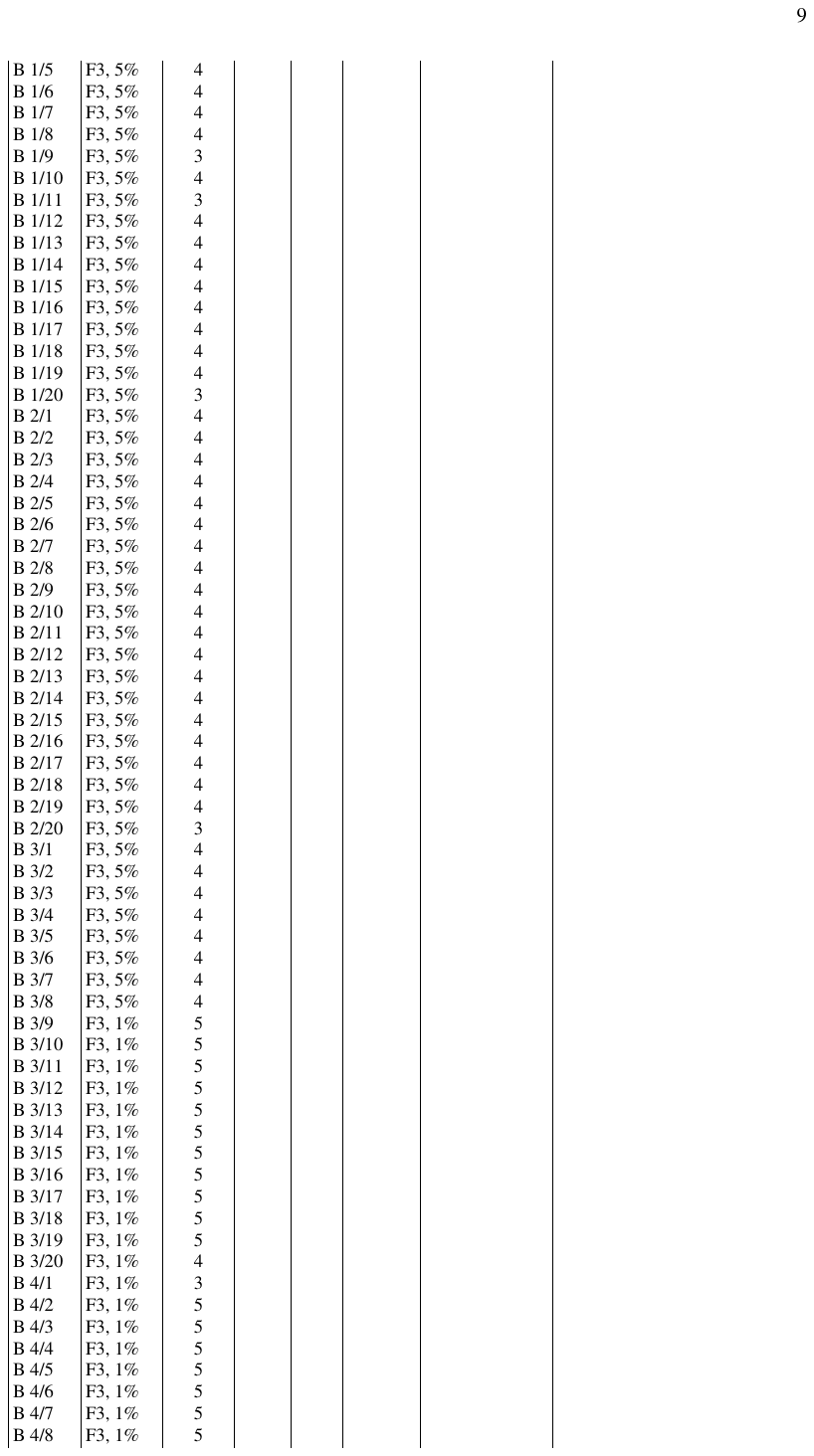}}
\clearpage

\hspace*{0cm}\fbox{\includegraphics[height=24cm,bb=50 60 570 800,clip] {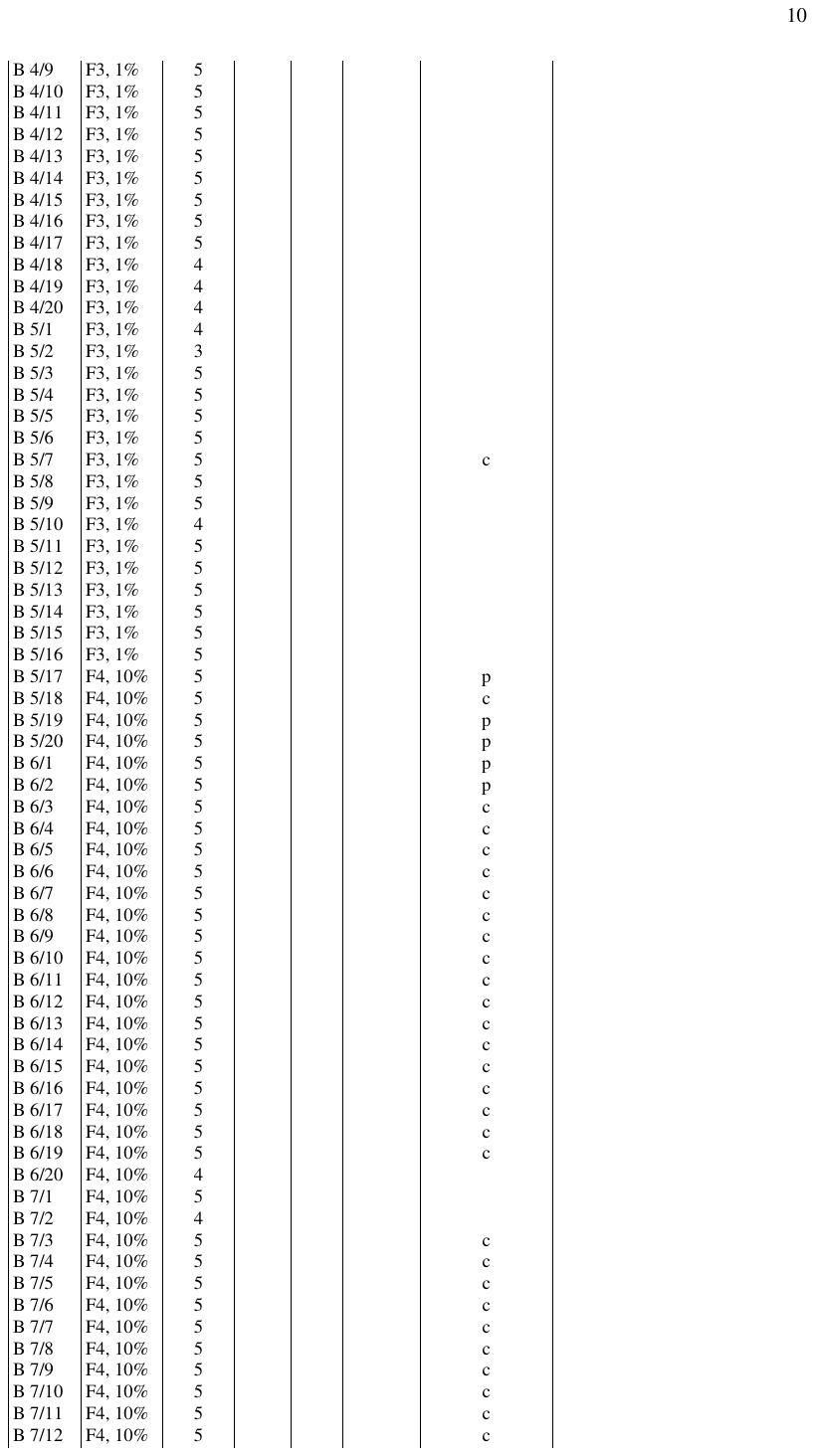}}
\clearpage

\hspace*{0cm}\fbox{\includegraphics[height=24cm,bb=50 60 570 800,clip] {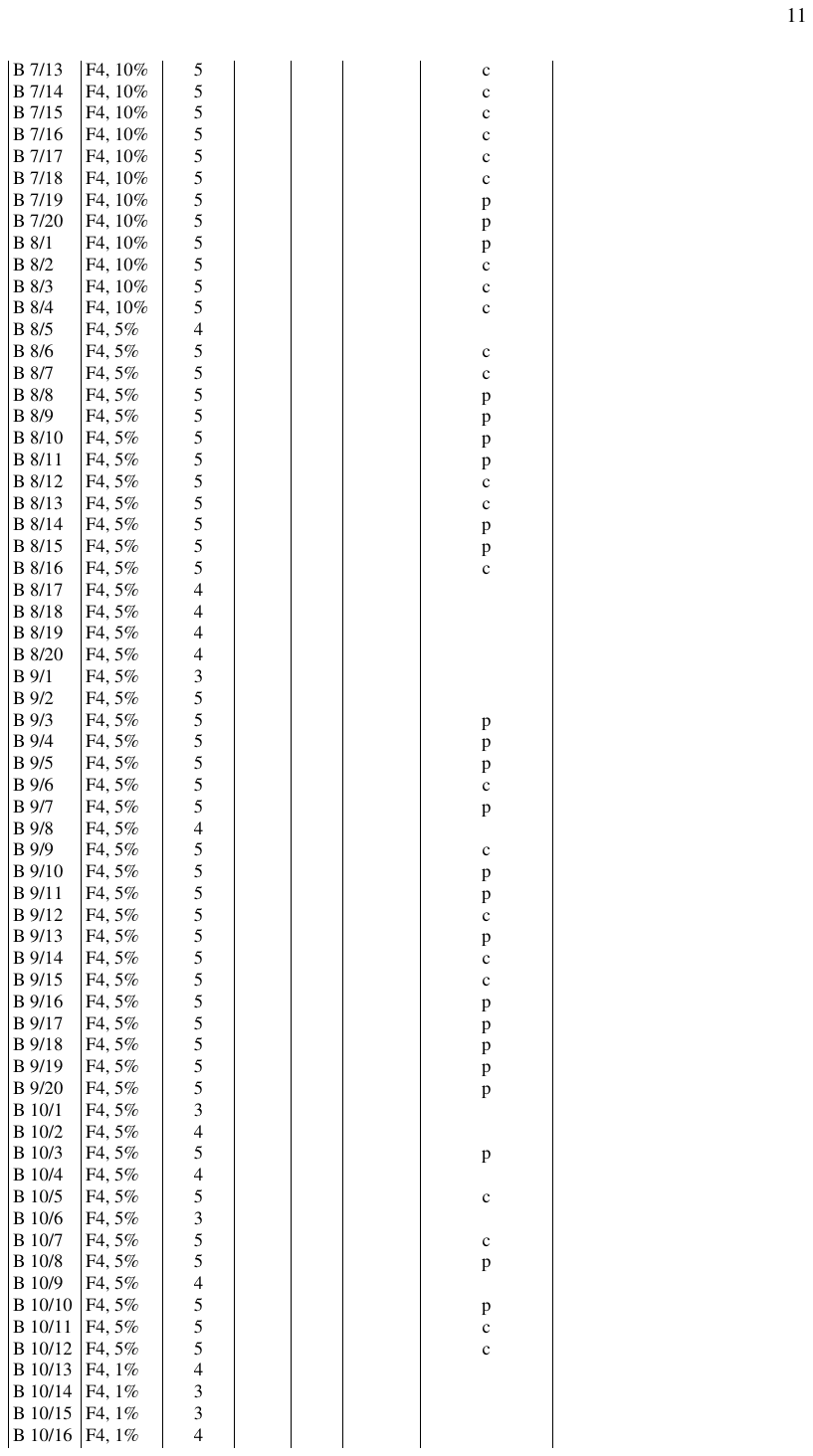}}
\clearpage

\hspace*{0cm}\fbox{\includegraphics[height=24cm,bb=50 60 570 800,clip] {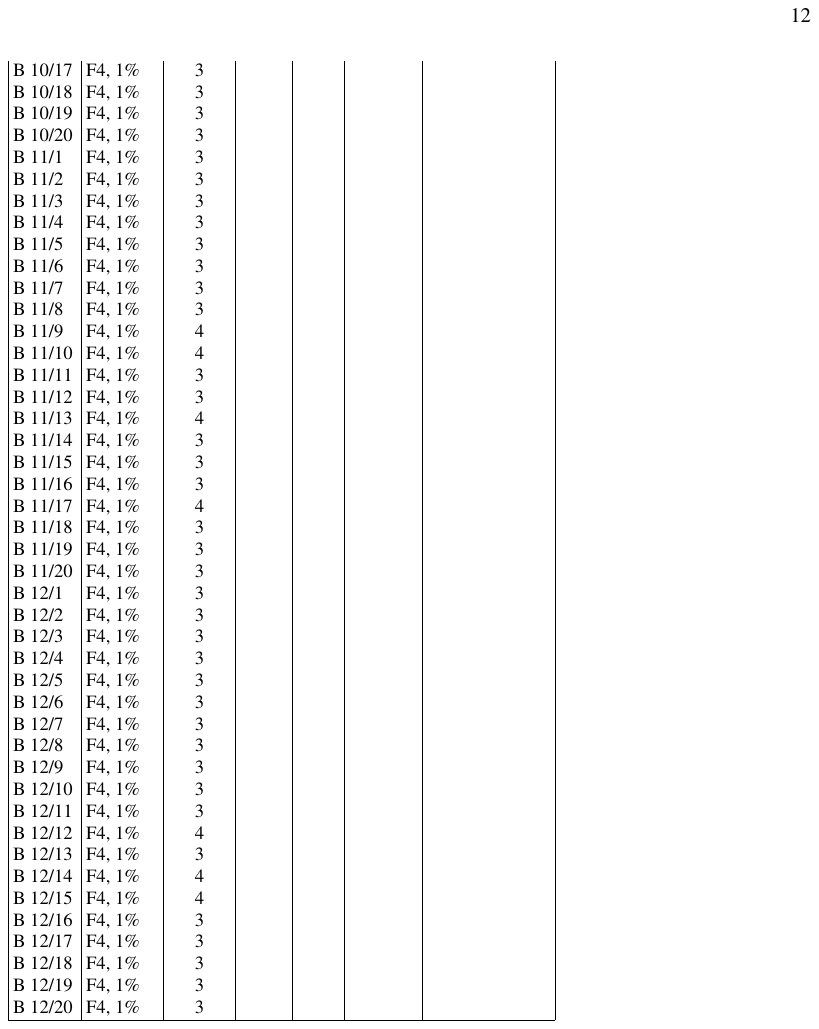}}
\clearpage

\end{document}